\documentstyle[aps,epsf,floats]{revtex}
\tighten
\newcommand{\ph}{{\rule{0mm}{3mm}}}
\newcommand{\slsh}[1]{\mbox{$\not\! #1$}}
\newcommand{\bm}[1]{\bbox{#1}}

\begin{document}
 
 
\draft
 
\title{
\begin{flushright}
\begin{minipage}{3 cm}
\small
hep-ph/9702281\\
NIKHEF 97-008\\ 
VUTH 97-3
\end{minipage}
\end{flushright}
Asymmetries in polarized hadron production\\
in $e^+e^-$ annihilation up to order $1/Q$}

\author{D. Boer$^1$, R. Jakob$^1$ and P.J. Mulders$^{1,2}$}
\address{\mbox{}\\
$^1$National Institute for Nuclear Physics and High--Energy
Physics (NIKHEF)\\	
P.O. Box 41882, NL-1009 DB Amsterdam, the Netherlands\\
\mbox{}\\
$^2$Department of Physics and Astronomy, Free University \\
De Boelelaan 1081, NL-1081 HV Amsterdam, the Netherlands
}

\maketitle
\begin{center}February, 1997\end{center}

\begin{abstract}
We present the results of the tree-level calculation of inclusive two-hadron 
production in electron-positron annihilation via 
one photon up to subleading
order in $1/Q$. We consider the situation where the two hadrons 
belong to different, back-to-back jets.
We include polarization of the produced hadrons and 
discuss azimuthal dependences of asymmetries. New asymmetries are found, in
particular there is a {\em leading\/} $\cos(2\phi)$ asymmetry, which is even 
present when 
hadron polarization is absent, since it arises solely due to the intrinsic  
transverse momenta of the quarks. 
\end{abstract}


\section{Introduction}

Three basic hard scattering processes in which the structure of hadrons
is studied with electroweak probes are (semi-)inclusive lepton-hadron 
scattering, the Drell-Yan process and inclusive hadron production in $e^+e^-$ 
annihilation. In this paper we will focus on the latter, where we restrict 
ourselves to photon exchange. The (timelike) photon momentum $q$ sets the 
scale $Q$, where $Q^2 \equiv q^2$, which is much larger than characteristic
hadronic scales. 

The inclusive lepton-hadron scattering, generally known as deep inelastic 
scattering (DIS), is the most studied process from the theoretical as well as 
from the experimental side. From the theoretical point of view DIS can be 
described by using the operator product expansion (OPE), within the context of
Quantum Chromodynamics (QCD). This allows to relate moments of structure 
functions to (Fourier transforms of) hadronic matrix elements of local 
operators. In the (QCD improved) parton model, i.e., to leading order in 
$1/Q$, the structure functions can be expressed as sums of so-called 
distribution functions (DF's). The OPE thus gives information on {\em moments}
of the DF's. However, one can also describe a 
scattering process directly in terms of the DF's themselves, which are 
(Fourier transforms of) hadronic matrix elements of non-local operators
\cite{Soper-77,Collins-Soper-82}. For 
treating the non-leading orders in $1/Q$, Ellis, Furmanski, Petronzio 
(EFP) \cite{Ellis-Furmanski-Petronzio-83} have developed a formalism for 
unpolarized DIS, using non-local operators, 
following ideas by Politzer \cite{Politzer-80}.
The extension to polarized DIS was done by Efremov and Teryaev 
\cite{Efremov-Teryaev-84}. At tree level, i.e., 
order $(\alpha_s)^0$, EFP have shown the equivalence of their formalism to 
the OPE approach for $(1/Q)^n$ power corrections. 

The non-local operators consist of quark and gluon fields. 
The quark (gluon) DF's are functions of the light-cone momentum fraction 
$x=p^+/P^+$ of a quark (gluon) with momentum $p$ in a hadron with 
momentum $P$. For multi-parton DF's, which show up beyond leading order, 
the functions depend on momentum fractions of several partons 
\cite{Politzer-80,Jaffe-83}. Often the number of partons can be reduced with
help of the QCD equations of motion.

The EFP approach was extended to other processes in analogy to the DIS 
description, in particular, to those ones where the OPE is not applicable
(for an alternative approach based on a non-local operator expansion see 
\cite{Balitsky-Braun-91}).
The Drell-Yan process has been studied to leading order by Ralston and Soper 
\cite{Ralston-Soper-79} and partially to subleading order in 
\cite{Jaffe-Ji-92,Qiu-Sterman-91,Tangerman-Mulders-95a}. Here the cross-section
is a sum of products of 
two DF's. Recently, the complete tree-level result up to order $1/Q$ for 
semi-inclusive polarized lepton-hadron scattering was published 
\cite{Mulders-Tangerman-96}. In this case one needs to include so-called 
fragmentation functions (FF's) \cite{Mueller-78,Collins-Soper-82} 
and the cross-section is a sum of products of a DF and a FF. 

The present paper focusses on the third process of interest: inclusive
two-hadron  
production in $e^+e^-$ annihilation up to order $1/Q$, where the two hadrons 
belong to different, back-to-back jets. The cross-section 
involves products of FF's, the number of which is larger than the number of 
DF's, due to the 
appearance of so-called time-reversal odd structures, which arise since 
time reversal does not give constraints in this case.
Interesting features like the Collins effect \cite{Collins-93b} show up, as do 
other new asymmetries.  

The DF's and FF's are essentially non-perturbative objects and must be 
determined by experiment or calculated, for instance with the help of models. 
The idea is 
that these functions are universal and once they are measured in one process 
they can lead to predictions in others. The DF's and FF's parametrize the 
structure of a hadron, e.g.\ the spin structure, and the leading order DF's 
and FF's have simple probabilistic interpretations. With the help of symmetry 
properties one can first of all determine the set of DF's and FF's in which 
a process can be expressed. By giving cross-sections and asymmetries in 
terms of these functions one can deduce how to measure and separate them. 

From the experimental side most well-known are the leading order unpolarized
and polarized DF's called $f_1$ and $g_1$, respectively, \cite{expDF's} 
and the FF $D_1$ \cite{expFF's}. 
For subleading order the DF $g_2$ has been measured \cite{SLAC}, 
but still with rather large errors. The experimental knowledge on FF's is 
much smaller than that on DF's. We obtain the 
most general expression for the cross-section in terms of as yet unknown FF's, 
in order to find out which other functions might be experimentally accessible
in the near future and where to look for them. Some of the unknown functions 
have been modelled \cite{Hoodbhoy}, which results 
can be used to estimate the magnitudes of asymmetries from our expressions. 

Some asymmetries, like the Collins effect, are leading effects, not suppressed 
by powers of $1/Q$, but one needs to include intrinsic transverse momentum 
in the DF's and FF's
\cite{Soper-77,Ralston-Soper-79,Collins-Soper-82}. 
Intrinsic transverse 
momentum plays a crucial role in two-hadron processes 
involving two soft, non-perturbative, parts, since the transverse 
momenta are linked by momentum conservation. 
Those DF's or FF's which, as functions of transverse momentum, would not 
contribute in one-hadron processes, will show up in these two-hadron 
processes. The idea that intrinsic transverse momentum always gives rise 
to suppression is incorrect, although it is the case in one-hadron processes.  
In the pioneering work \cite{Cahn-78,Berger-79,Berger-80} on azimuthal 
dependences due to 
intrinsic transverse momentum, all effects are found to be suppressed by at
least $1/Q$.  

On the formal theoretical side there remains the open issue of how to proof
factorization for the process under 
consideration. For the case of back-to-back jets there exists a proof, but
no higher twist effects are included \cite{Collins-Soper-81}. We do not 
expect polarization to be a problem for factorization (see \cite{Collins-93a}).
For the case of the Drell-Yan process, which is very similar to the 
process under consideration, arguments have been given why factorization 
holds for the first non-leading power corrections 
\cite{Qiu-Sterman-91b,Qiu-Sterman-90}, which is consistent with its known 
failure at order $1/Q^4$ \cite{Doria-80,Dilieto-81,Basu-84}. We expect 
factorization to hold also for the case at hand.     

We will 
not concern ourselves with these problems here, even though 
factorization is needed to ensure universality of the FF's, and 
restrict the discussion to tree-level.
It represents the extension of the 
naive parton model to subleading order and shows the dominant structures to 
be expected in the cross-section, although QCD corrections, such as Sudakov 
effects \cite{Dokshitzer-80}, may affect the magnitude of the asymmetries. 
At tree-level the 
only QCD input at order $1/Q$ (apart from the Feynman rules) is the use of the
equations of motion which ensures the electromagnetic gauge invariance. 

The outline of this paper is as follows. In Section 2 we present the formalism
of the $e^+e^-$ annihilation process, with emphasis on the kinematics.
Section 3 contains the analysis of the soft parts of the process, 
in particular
the fragmentation functions are studied. This is followed by the details of 
the complete tree-level calculation of the hadron tensor up to subleading 
order in Section 4, the result of which is given
in Appendix B. In the three following sections we investigate special 
cases, which give more insight than the full result 
and are useful from a practical point of view. 
In Section 5 we discuss the result after
integration over the transverse momentum of the photon. In Section 6 we study
the differential cross-section, i.e., not integrated over transverse photon 
momentum, but restricted to leading order and the case
where only one hadron is polarized (the case of two polarized hadrons is given
in Appendix C). In Section 7 this is compared with the 
integrated cross-section weighted with factors of the transverse momentum of 
the photon. Finally, the results are summarized in Section~8. 

\section{Kinematics}

We consider $e^-+e^+$ $\rightarrow$ hadrons, where the two leptons with
momenta $l$ and $l^\prime$ annihilate into a photon with momentum $q =
l + l^\prime$, which is timelike with $q^2 \equiv Q^2 \rightarrow \infty$.
Denoting the momentum of outgoing hadrons by $P_h$ ($h$ = 1, 2, \ldots) we
use invariants $z_h$ = $2P_h\cdot q/Q^2$. The momenta can also be considered
as jet momenta.  We will consider the general case of polarized leptons with
helicities $\pm \lambda_e$ and production of hadrons of which
the spin states are characterized by a spin vector $S_h$ ($h$ = 1, 2, \ldots),
satisfying $S_h^2 = -1$ and $P_h\cdot S_h = 0$. In this way we can treat the
case of 
unpolarized final states or final state hadrons with spin-0 and spin-1/2.
We will work in the limit where  $Q^2$ and $P_h\cdot q$ are large, keeping 
the ratios $z_h$ finite. 

The square of the amplitude can be split into a purely leptonic and a
purely hadronic part, 
\begin{equation}
|{\cal M}|^2 = \frac{e^4}{Q^4} L_{\mu\nu} H^{\mu\nu} ,
\end{equation}
with the helicity-conserving lepton tensor (neglecting the lepton masses) 
given by
\begin{equation} \label{leptten2}
L_{\mu\nu} (l, l^\prime; \lambda_e)
=  2 l_\mu l^\prime_\nu
+ 2 l_\nu l^\prime_\mu - Q^2 g_{\mu\nu}
+2i\lambda_e \,\epsilon_{\mu\nu\rho\sigma} l^\rho l^{\prime \sigma} .
\end{equation}
For the case of two observed hadrons in the final state, the product of 
hadronic current matrix elements is written as
\begin{equation}
H_{\mu\nu}(P_X;  P_1 S_1; P_2 S_2)
= \langle 0 |J_\mu (0)|P_X; P_1 S_1; P_2 S_2 \rangle
\langle P_X; P_1 S_1; P_2 S_2 |J_\nu (0)| 0 \rangle ,
\end{equation}
where a summation over spins of the unobserved {\em out}-state is understood.
The cross-section (including a factor 1/2 from averaging over incoming
polarizations) is given by: for {\em 2-particle inclusive} $e^+e^-$ 
annihilation
\begin{equation}
\frac{P_1^0 \,P_2^0\,\,d\sigma^{(e^+e^-)}}{d^3P_1\,d^3P_2}
=\frac{\alpha^2}{4\,Q^6} L_{\mu\nu}{\cal W}^{\mu\nu},
\label{cross2}
\end{equation}
with
\begin{equation}
{\cal W}_{\mu\nu}( q;  P_1 S_1; P_2 S_2 ) =
\frac{1}{(2\pi)^4} \int \frac{d^3 P_X}{(2\pi)^3 2P_X^0}
(2\pi)^4 \delta^4 (q-P_X - P_1 - P_2)
H_{\mu\nu}(P_X;  P_1 S_1; P_2 S_2),
\label{hadrten2}
\end{equation}
for {\em 1-particle inclusive} $e^+e^-$ annihilation
\begin{equation}
P_h^0\,\frac{d\sigma}{d^3 P_h} = \frac{\alpha^2}{2\,Q^6}\,L_{\mu\nu}
W^{\mu \nu},
\end{equation}
with
\begin{equation} 
W_{\mu\nu}( q;  P_h S_h) =
\frac{1}{(2\pi)}
\int \frac{d^3 P_X}{(2\pi)^3 2P_X^0}
(2\pi)^4 \delta^4 (q-P_X - P_h )
\langle 0 |J_\mu (0)|P_X; P_h S_h \rangle
\langle P_X; P_h S_h |J_\nu (0)| 0 \rangle ,
\end{equation}
and for the {\em totally inclusive} annihilation cross-section the well-known 
result
\begin{equation}
\sigma(e^+e^- \rightarrow hadrons) =
\frac{4\pi^2\,\alpha^2}{Q^6}\,L_{\mu\nu} R^{\mu \nu},
\end{equation}
with the tensor $R_{\mu\nu}$ given by
\begin{eqnarray}
R_{\mu\nu}(q) & = &
\int \frac{d^3 P_X}{(2\pi)^3 2P_X^0}
(2\pi)^4 \delta^4 (q-P_X)
\langle 0 |J_\mu (0)|P_X\rangle\langle P_X |J_\nu (0)| 0 \rangle \nonumber
\\ & = & \int d^4 x\ e^{iq\cdot x}\,
\langle 0 |[J_\mu (x),J_\nu (0)]| 0 \rangle  .
\end{eqnarray}
Recall that the totally inclusive cross-section is directly related to the 
vacuum polarization. Also note that the totally inclusive process is 
short-distance dominated, whereas the 1-particle inclusive case is light-cone 
dominated, but only in the former the OPE can be applied.

In order to expand the lepton and hadron tensors in terms of independent
Lorentz structures, it is convenient to
work with vectors orthogonal to $q$. A normalized timelike vector is defined 
by $q$ and
a normalized spacelike vector is defined by $\tilde P^\mu$ = 
$P^\mu - (P\cdot q/q^2)\,q^\mu$ for one of the outgoing momenta, say $P_2$,
\begin{eqnarray}
\hat t^\mu& \equiv & \frac{q^\mu}{Q},\\
\hat z^\mu & \equiv &
\frac{Q}{P_2\cdot q}\,\tilde P^\mu_2
\ =\   2\,\frac{P_2^\mu}{z_2 Q} - \frac{q^\mu}{Q}. 
\end{eqnarray}
\begin{figure}[t]
\begin{center}
\leavevmode \epsfxsize=10cm \epsfbox{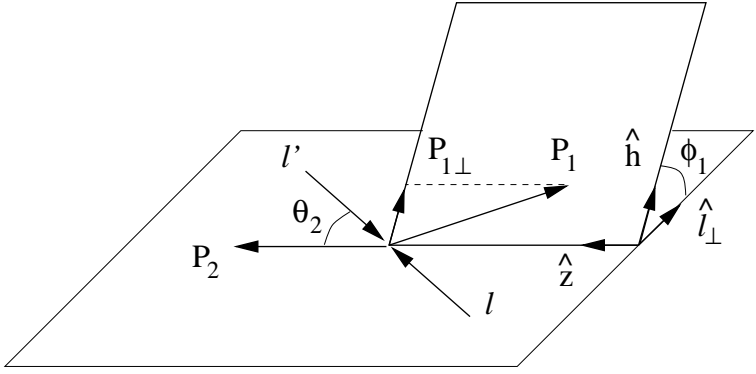}
\vspace{0.2 cm}
\caption{\label{fig:kinann} Kinematics of the annihilation process in
the lepton center of mass frame for a back-to-back jet situation.
$P_2$ is the momentum of a jet or of a fast hadron in a jet, $P_1$ is
the momentum of a hadron belonging to the other jet.}
\end{center}
\end{figure}
Note that we have neglected $1/Q^2$ corrections, as we will do throughout the
paper. Such 
corrections arise among others from the hadron masses $M_h$, so-called target 
mass corrections or kinematic power corrections. 

Vectors orthogonal to $\hat z$ and $\hat t$ are obtained with help of 
the tensors
\begin{eqnarray}
&& g_{\perp}^{\mu\nu}
\equiv  g^{\mu\nu} -\hat t^\mu \hat t^\nu +
\hat z^\mu \hat z^\nu, \\
&& \epsilon_\perp^{\mu \nu} \equiv
-\epsilon^{\mu \nu \rho \sigma} \hat t_\rho \hat z_\sigma
\ =\ \frac{1}{(P_2\cdot q)}\,\epsilon^{\mu \nu \rho
\sigma} P_{2\,\rho}q_\sigma.
\end{eqnarray}
For instance, using the other hadronic momentum $P_1$ 
, one obtains $P_{1\perp}^\mu$ = $g_\perp^{\mu\nu}\,P_{1\nu}$ (see Fig.\ 
\ref{fig:kinann}). We define 
the normalized vector $\hat h^\mu$ = $P_{1\perp}^\mu/\vert \bm P_{1\perp}
\vert$ and the second orthogonal direction is given by 
$\epsilon_\perp^{\mu \nu} \hat h_\nu$. 
We use boldface vectors to denote the two-dimensional Euclidean part
of a four-vector, such that $P_{1\perp}\cdot P_{1\perp}=-\bm P_{1\perp}\cdot 
\bm P_{1\perp}$. 
In the calculation of the hadron tensor it will be convenient to define 
lightlike directions using the
hadronic (or jet) momenta. Consider two hadronic momenta $P_1$ and $P_2$ 
not belonging to one jet (i.e., their dot product $P_1\cdot P_2$ is of
order $Q^2$). The momenta can then be parametrized using dimensionless 
lightlike vectors
$n_+$ and $n_-$ satisfying $n_+^2 = n_-^2 = 0$ and $n_+\cdot n_-$ = 1,
\begin{eqnarray}
&& P_1^\mu \equiv \frac{\zeta_1\tilde Q}{\sqrt{2}}\,n_-^\mu
+ \frac{M_1^2}{\zeta_1\tilde Q\sqrt{2}}\,n_+^\mu,\\
&& P_2^\mu \equiv \frac{M_2^2}{\zeta_2\tilde Q\sqrt{2}}\,n_-^\mu + 
\frac{\zeta_2\tilde Q}{\sqrt{2}}\,n_+^\mu,\\
&&q^\mu \equiv \frac{\tilde Q}{\sqrt{2}}\,n_-^\mu
+ \frac{\tilde Q}{\sqrt{2}}\,n_+^\mu + q_T^\mu,
\end{eqnarray}
where $\tilde Q^2$ = $Q^2 + Q_T^2$ with $q_T^2 \equiv -Q_T^2$. We will use the 
notation $p^- = p \cdot n_+$ and $p^+ = p \cdot n_-$ for a generic momentum 
$p$. For the case
of two back-to-back jets $Q_T^2 \ll Q^2$ and up to $Q_T^2/Q^2$, which we
neglect, one has $\tilde Q = Q$, $\zeta_1 = z_1$ and $\zeta_2 = z_2$. If 
momentum $P_2$ is used to define the vector $\hat z^\mu$, then
\begin{equation}
P_{1\perp}^\mu = - z_1\,q_T^\mu.
\end{equation}
Vectors transverse to $n_+$ and $n_-$ one obtains using the tensors
\begin{eqnarray}
&&g^{\mu\nu}_T \ \equiv \ g^{\mu\nu}
- n_+^{\,\{\mu} n_-^{\nu\}}, \\
&&\epsilon^{\mu\nu}_T \ \equiv
\ \epsilon^{\mu\nu\rho\sigma} n_{+\rho}n_{-\sigma},
\end{eqnarray}
where the brackets around the indices indicate symmetrization. Note that 
these {\em transverse} tensors are not identical to
the {\em perpendicular} ones defined above if the transverse momentum
of the outgoing hadron does not vanish.
The lightlike directions, however, can easily be expressed in $\hat t$,
$\hat z$ and a perpendicular vector,
\begin{eqnarray}
n_+^\mu & = & \frac{1}{\sqrt{2}} \left[ \hat t^\mu + \hat z^\mu \right],
\label{transverse1} \\
n_-^\mu & = & \frac{1}{\sqrt{2}} \left[ \hat t^\mu - \hat z^\mu
- 2\,\frac{q_T^\mu}{Q} \right] =
\frac{1}{\sqrt{2}} \left[ \hat t^\mu - \hat z^\mu
- 2\,\frac{Q_T^{}}{Q}\,\hat h^\mu \right], 
\label{transverse2}
\end{eqnarray}
showing that the differences are of order $1/Q$. Especially for the treatment
of azimuthal asymmetries, it is important to keep track of these differences.

In summary, we use two sets of basis vectors, the first set constructed from  
the photon momentum ($q$) and one of the hadron momenta ($P_2$), the second 
set from the two hadron momenta ($P_1$ and $P_2$). The 
respective frames where the momenta $q$ and $P_2$, or $P_1$ and $P_2$, are
collinear are the natural ones connected to these two sets. 
In the first $P_1$ has a perpendicular
component $P_{1\perp}$, in the second $q$ has a transverse component
$q_T^{}$. One can get from one frame to the other via a Lorentz
transformation that leaves the minus components unchanged 
\cite{Levelt-Mulders-94}.

\section{Correlation functions}

In this section we discuss the relevant `soft' hadronic matrix elements that 
appear
in the diagrammatic expansion of a hard scattering amplitude. Assuming the two
hadrons to belong to two different jets we encounter two types of
soft parts in the process
under consideration: one describes the
fragmentation of a quark into a hadron plus a remainder which is not detected
and the other describes the similar fragmentation for an antiquark. Up to
order $1/Q$ the quark fragmentation is described with help of two types of 
correlation
functions: the quark-quark correlation function $\Delta (P_1,S_1;k)$ 
\cite{Collins-Soper-82}
and the
quark-gluon-quark correlation function $\Delta_A^\alpha (P_1,S_1;k,k_1)$ 
(Fig.~\ref{corrfns}):
\begin{figure}[htb]
\begin{center}
\leavevmode \epsfxsize=10cm \epsfbox{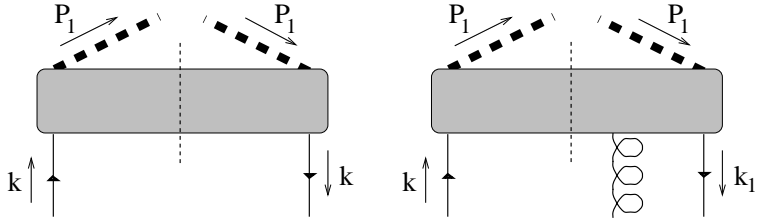}
\caption{\label{corrfns} Correlation functions $\Delta$ and $\Delta_A$.}
\end{center}
\end{figure}
\begin{eqnarray}
\Delta_{ij}(P_1,S_1;k) & = & \sum_X \frac{1}{(2\pi)^4}\int d^4x\ e^{ik\cdot 
x}
\,
\langle 0 \vert \psi_i(x) \vert P_1,S_1; X \rangle
\langle P_1, S_1; X \vert \overline \psi_j(0) \vert 0 \rangle, \\
 \Delta^\alpha_{A\, ij}(P_1,S_1; k,k_1) &=& \sum_X \frac{1}{(2\pi)^4}
 \int d^4x\,d^4y\ e^{ i\,k\cdot y + i\,k_1\cdot(x-y) } \langle 0 \vert
\psi_i(x)\, g A_T^\alpha (y)\, \vert P_1,S_1; X \rangle
\langle P_1, S_1; X \vert \,\overline \psi_j(0)\vert 0\rangle,
\end{eqnarray}
where $k, k_1$ are the quark momenta and an averaging over color indices 
is understood. If one chooses the gauge $A^-=0$ only a transverse gluon is  
relevant. In fact, in a calculation up to subleading
order, we only encounter the partly integrated correlation functions
$\int dk^+\, \Delta (P_1,S_1;k)$ and $\int dk^+ d^4k_1 \,
\Delta_A^\alpha (P_1,S_1;k,k_1)$, which are functions of $k^-$ and 
$\bm{k}_T^{}$ only.
Note that the definition of $\Delta_A^\alpha$ includes one power of the
strong coupling constant $g$. 

The above matrix elements as functions of invariants are assumed to vanish
sufficiently fast above a
characteristic hadronic scale, which is much smaller than $Q^2$.
This means that in the above matrix elements $k^2, k\cdot P_1 \ll Q^2$. 
Hence, we make the following Sudakov decomposition for the quark momentum $k$:
\begin{equation}
k \equiv \frac{z_{1} Q}{z\sqrt{2}}\,n_-
+ \frac{z (k^2 + \bm{k}_T^2)}{z_{1} Q \sqrt{2}}\,n_+ + k_T^{}
 \approx  \frac{1}{z} P_1 + k_T^{}.
\end{equation}
Similarly, we decompose the spin vector $S_1$:
\begin{equation}
S_1 \equiv \frac{\lambda_1 z_{1} Q}{M_1  \sqrt{2}}\,n_-
-\frac{\lambda_1 M_1}{z_{1} Q \sqrt{2}}\,n_+ + S_{1T}^{}
\approx  \frac{\lambda_1}{M_1} P_1 + S_{1T}^{},
\end{equation}
with for a pure state $\lambda_1^2+\bm{S}_{1T}^2=1$. 
In the approximations the $+$ components ($\propto 
1/Q$) are neglected, as these are irrelevant compared to the $+$ components 
of the momenta in the hard part ($\propto Q$). 

The Dirac structure of the quark-quark correlation function can be expanded in 
a number of amplitudes,
i.e., functions of invariants built up from the quark and hadron momenta, 
constrained by hermiticity and parity 
\cite{Ralston-Soper-79,Mulders-Tangerman-96}.
Here we directly integrate the correlation function over $k^+$, which up to 
order $1/Q$ can be parametrized as follows:
\begin{eqnarray}
&&\left. \frac{1}{4z} \int dk^+\ \Delta(P_1,S_1;k)
\right|_{k^- = P_1^-/z,\ \bm{k}_{\scriptscriptstyle T}} =
\frac{M_1}{4P_1^-} \Biggl\{
E\, {\bf 1}
+ D_1\,\frac{\slsh{\!P_1}}{M_1}
+ D_{1T}^\perp\, \frac{\epsilon_{\mu \nu \rho \sigma}
\gamma^\mu P_1^\nu k_T^\rho S_{1T}^\sigma}{M_1^2}
+ D^\perp\,\frac{\slsh{k_T^{}}}{M_1}
\nonumber \\[3mm]
&& \qquad + D_T\,\epsilon_{\mu\nu\rho\sigma}
n_+^\mu n_-^\nu\gamma^\rho S_{1T}^\sigma
+ \lambda\,D_L^\perp\,\frac{\epsilon_{\mu\nu\rho\sigma}
n_+^\mu n_-^\nu\gamma^\rho k_T^\sigma}{M_1}
- E_{s} \, i\gamma_5
- G_{1s}\,\frac{\slsh{\!P_1} \gamma_5}{M_1}
- G_{T}^\prime\,\slsh{S_{1T}^{}} \gamma_5
\nonumber \\[3mm] & &
\qquad - G_{s}^\perp\, \frac{\slsh{k_T^{}} \gamma_5}{M_1}
- H_{1T}\,\frac{i\sigma_{\mu\nu}\gamma_5\,S_{1T}^\mu P_1^\nu}{M_1}
- H_{1s}^\perp\,
\frac{i\sigma_{\mu\nu}\gamma_5\, k_T^\mu P_1^\nu}{M_1^2}
- H_T^\perp\,
\frac{i\sigma_{\mu\nu}\gamma_5\, S_{1T}^{\mu} k_T^\nu}{M_1}
- H_{s}\,i\sigma_{\mu\nu}\gamma_5\, n_{-}^{\mu} n_{+ }^{\nu}
\nonumber \\[3mm] & &
\qquad + H_{1}^\perp\,\frac{\sigma_{\mu \nu} k_T^\mu P^\nu}{M_1^2}
+ H\,\sigma_{\mu \nu} n_-^\mu n_+^\nu
\Biggr\},
\label{Deltaexp}
\end{eqnarray}
where the shorthand notation $G_{1s}$ stands for the combination
\begin{equation}
G_{1s}(z,\bm{k}_T^{}) = \lambda_1\,G_{1L}
+ G_{1T}\,\frac{(\bm{k}_T^{}\cdot \bm{S}_{1T}^{})}{M_1},
\end{equation}
etc. 
The functions $E, D_1, \ldots$ in Eq.\ (\ref{Deltaexp}) and $G_{1L}, G_{1T},
\ldots$ in $G_{1s}, \ldots$ are fragmentation functions. One wants to express
the fragmentation functions in terms of the hadron momentum, hence, the
arguments of
the fragmentation functions are chosen to be the lightcone (momentum)
fraction $z = P_1^-/k^-$ of the produced hadron with respect to the
fragmenting quark and $\bm{k}_T^\prime \equiv -z\bm{k}_T^{}$, which is the 
transverse 
momentum of the hadron in a frame where the quark has no transverse momentum.
In order to switch from
quark to hadron transverse momentum a Lorentz transformation leaving $k^-$
and $P_1^-$ unchanged needs to be performed. The fragmentation
functions are real and in fact, depend on $z$ and $\bm{k}_T^\prime{}^2$ only.

Inverting the above expression, the fragmentation functions appear in specific 
Dirac projections of the correlation functions, integrated over $k^+$:
\begin{eqnarray}
\Delta^{[\Gamma]}(z,\bm{k}_T^{}) & \equiv &
\left. \frac{1}{4z}\int dk^+\ \text{Tr}(\Delta\,\Gamma)\right|_{k^- =
P_1^-/z,\ \bm{k}_{\scriptscriptstyle T}} \nonumber \\
& = & \left. \sum_X \int \frac{dx^+d^2\bm{x}_T^{}}{4z\,(2\pi)^3} \
e^{ik\cdot x} \, \text{Tr}  \langle 0 \vert \psi (x) \vert P_1,S_1; X \rangle
\langle P_1, S_1; X \vert
 \overline \psi(0) \Gamma \vert 0 \rangle \right|_{x^- = 0} ,
\label{projections}
\end{eqnarray}
for which we can distinguish the leading fragmentation functions:
\begin{eqnarray} 
& &\Delta^{[\gamma^-]}(z,\bm{k}_T^{}) =
D_1(z,\bm{k}_T^\prime{}^2) + \frac{\epsilon_T^{ij}k_{Ti}^{} 
S_{1Tj}^{}}{M_1}
\,D_{1T}^\perp(z,\bm{k}_T^\prime{}^2),
\label{projectionsL}
\\ & &\Delta^{[\gamma^-\gamma_5]}(z,\bm{k}_T^{}) =
G_{1s}(z,\bm{k}_T^{}),
\\ & & \Delta^{[i \sigma^{i-} \gamma_5]}(z,\bm{k}_T^{}) =
S_{1T}^i\,H_{1T}(z,\bm{k}_T^\prime{}^2)
+ \frac{k_T^i}{M_1}\,H_{1s}^\perp(z,\bm{k}_T^{})
+ \frac{\epsilon_T^{ij} k_{Tj}^{}}{M_1}\,H_1^\perp(z,\bm{k}_T^\prime{}^2);
\end{eqnarray}
furthermore we obtain subleading
projections ($i,j$ are transverse indices):
\begin{eqnarray}
& &\Delta^{[{\bf 1}]}(z,\bm{k}_T^{}) =
 \frac{M_1}{P_1^-}\,E(z,\bm{k}_T^\prime{}^2) ,
\\ & & \Delta^{[\gamma^i]}(z,\bm{k}_T^{}) =
\frac{k_T^i}{P_1^-}\,D^\perp(z,\bm{k}_T^\prime{}^2)
+ \frac{\lambda_1\,\epsilon_T^{ij}k_{Tj}^{}}{P_1^-}\,D_L^\perp(z, 
\bm{k}_T^\prime{}^2)
+ \frac{M_1\,\epsilon_T^{ij}S_{1T\,j}^{}}{P_1^-}\,D_T(z, 
\bm{k}_T^\prime{}^2),
\\ & &\Delta^{[i\gamma_5]}(z,\bm{k}_T^{}) =
\frac{M_1}{P_1^-} \,E_{s}(z,\bm{k}_T^{}),
\\  & &\Delta^{[\gamma^i \gamma_5]}(z,\bm{k}_T^{}) =
\frac{M_1\,S_{1T}^i}{P_1^-} \, G_{T}^\prime(z,
\bm{k}_T^\prime{}^2)
+ \frac{k_T^i}{P_1^-}\,G_{s}^\perp(z,\bm{k}_T^{}),
\\ & & \Delta^{[ i \sigma^{ij} \gamma_5]}(z,\bm{k}_T^{}) =
\frac{S_{1T}^i k_T^j-k_T^i S_{1T}^j}{P_1^-}\,H_T^\perp(z,
\bm{k}_T^\prime{}^2)
+ \frac{M_1\,\epsilon_T^{ij}}{P_1^-}\,H(z,
\bm{k}_T^\prime{}^2),
\\ & & \Delta^{[i\sigma^{-+}\gamma_5]}(z,\bm{k}_T^{}) =
\frac{M_1}{P_1^-} \,H_{s}(z,\bm{k}_T^{}).
\label{projectionsSL}
\end{eqnarray}

We identified leading and subleading functions, which in principle start 
contributing at order $1$ and $1/Q$, respectively.
The order at which a function first can contribute depends on the power of
$M_1/P_1^-$ in front of the function as it appears in the projections.
Each factor $M_1/P_1^-$ leads to a suppression with a power of $M_1/Q$ in
cross-sections. We will refer to
the function multiplying a power $(M_1/P_1^-)^{t-2}$ as being of `twist' $t$.
We note that this notion of twist, in analogy to the 
$\bm{k}_T^{}$ integrated case
\cite{Jaffe-96}, is related but not equal
to the one used for local operators in the OPE. 

The naming scheme is as follows. All functions obtained after tracing with a
scalar (${\bf 1}$) or pseudoscalar ($i\gamma_5$) Dirac matrix are given the
name
$E_{..}$, those traced with a vector matrix ($\gamma^{\mu}$) are given the name
$D_{..}$, those traced with an axial vector matrix ($\gamma^\mu \gamma_5$) are
 given the name $G_{..}$ and, finally, those traced with the second rank
tensor ($i\sigma^{\mu\nu}\gamma_5$) are given the name $H_{..}$. A subscript
1 is given to the leading functions, subscripts $L$ or $T$ refer to the
connection with
the hadron spin being longitudinal or transverse and a superscript $\perp$
signals the explicit presence of transverse momenta with a non-contracted
index. In the literature sometimes the fragmentation functions are denoted
by lower-case names, but supplemented by a hat ($\hat{e}, \hat{g}, \hat{h}$),
with the one exception that $D$ is named $\hat{f}$. 
We note that after integration over $\bm k_T^{}$ several functions
disappear. In the case of $\Delta^{[i\sigma^{i-}\gamma_5]}$ and
$\Delta^{[\gamma^{i}\gamma_5]}$ specific combinations remain, 
namely $H_1$ $\equiv$ $H_{1T} + (\bm k_T^2/2M_1^2)\,H_{1T}^\perp$ and 
$G_T$ $\equiv$ $G_T^\prime + (\bm k_T^2/2M_1^2)\,G_T^\perp$, respectively.

The
choice of factors in the definition of fragmentation functions is such that
$\int dz\,d^2\bm k_T^\prime\,D_1(z,\bm k_T^\prime) = N_h$, where $N_h$ is 
the number of produced hadrons. 
The twist-two fragmentation functions have natural interpretations as decay
functions. The projection 
$\Delta^{[\gamma^-]}$ is (after proper normalizing) the probability of a quark
to produce a spin-1/2
hadron in a specific spin state, 
$\Delta^{[\gamma^-\gamma_5]}$ is the difference of the probabilities for a
chirally right and chirally left quark to produce such a hadron, while
$\Delta^{[i\sigma^{i-}\gamma_5]}$ is the difference of opposite transverse 
spin states (along direction i) of a quark to produce such a hadron.  

Note that the decay probability for an unpolarized quark with non-zero 
transverse momentum can lead to a transverse polarization in the production 
of spin-1/2 particles. This polarization is orthogonal to the quark transverse
momentum and the probability is given by the function $D_{1T}^{\perp}$. 
In the same way, oppositely transversely polarized quarks with
non-zero transverse momentum can produce unpolarized hadrons or 
spinless particles, with different probabilities. This difference is described
by the function $H_1^\perp$, which is the one appearing in the so-called
Collins effect \cite{Collins-93b}, which shows up as a single transverse 
spin asymmetry
in semi-inclusive DIS, and arises due to intrinsic transverse momentum. 

The functions $D_{1T}^{\perp}$ and $H_1^\perp$ are examples of what are 
generally called `time-reversal odd' functions. 
This somewhat misleading terminology refers to the behavior of 
the functions under the so-called {\em naive} time-reversal operation $T_N$ 
\cite{DeRujula-71}, which acts as follows on the correlation 
functions:
\begin{equation}
\Delta(P_1,S_1;k) \stackrel{T_N}{\longrightarrow} 
(\gamma_5 C \,\Delta(\bar P_1,\bar S_1; \bar k) \, C^\dagger \gamma_5)^\ast
\end{equation}
where $\bar k$ = $(k^0,-\bm{k})$, etc. If $T_N$ invariance would apply, 
the functions $D_{1T}^{\perp}$, $H_{1}^{\perp}$, $D_{L}^{\perp}$, $D_T$, 
$E_L$, $E_T$ and $H$ would be purely
imaginary. On the other hand, hermiticity requires the functions to be real,
so these functions should then vanish. 

The operation $T_N$ differs from the
actual time-reversal operation $T$ in that the former does not transforms 
{\em in} into {\em out}-states and vice versa. 
Due to final state interactions, the
{\em out}-state $\vert P_1,S_1; X \rangle$ in $\Delta(P_1,S_1;k)$ is not a 
plane
wave state and thus, is not simply related to an {\em in}-state. 
Therefore, one has 
$T_N\neq T$ and since $T$ itself does not pose any constraints on the 
functions, they need not vanish. 

In the analogous case of distribution functions, which are derived from matrix
elements with plane wave states, $T=T_N$ and therefore there are no 
`time-reversal odd' distribution functions.

The $H_{..}$ and $E$ functions (but not $E_s$) are called chiral-odd functions 
because they are non-diagonal in the chirality basis, so they arise either 
accompanied by
a quark mass term or by another chiral-odd function, such that the product is
again chiral-even \cite{Artru,Jaffe-Ji-92}.

The quark-gluon-quark correlation functions can be expressed in terms of the
quark-quark correlation functions with help of the classical equations of
motion (e.o.m.). These can be used inside hadronic matrix 
elements \cite{Politzer-80}. If we again define 
Dirac projections:
\begin{eqnarray}
\Delta_A^{\alpha [\Gamma ]}(z,\bm{k}_T^{}) & = &
\left. \frac{1}{4z} \int dk^+ \, d^4 k_1 \text{Tr} \left( 
\Delta_A^\alpha \Gamma 
\right)
\right|_{k^- \,=\,P_1^-/z,\ \bm{k}_{\scriptscriptstyle T}} \nonumber \\ & = &
\left. \sum_X \int \frac{dx^+ d^2x_\perp}{4z\,(2\pi)^3} \ e^{i\,k\cdot x}\,
\text{Tr}\,\langle 0 \vert \psi(x)\, g A_T^\alpha (x)\,
\vert P_1,S_1; X \rangle \langle P_1, S_1; X \vert\,
\overline \psi(0) \Gamma \vert 0 \rangle \right|_{x^- \,=\, 0} ,
\end{eqnarray}
we find as a consequence of the e.o.m.:
\begin{eqnarray}
\Delta_{A\,\alpha}^{\ [\sigma^{\alpha -} ]} =
-\epsilon_T^{\alpha \beta}\,\Delta_A^{\alpha [i\sigma^{\beta -}\gamma_5]}
&=& \frac{M_1}{z}\left(\tilde{H} + i\,\tilde{E} \right)
- \epsilon_T^{ij}\,k_{Ti}^{} S_{1T\,j}^{} \, 
\left(\frac{1}{z}\tilde{H}_T^\perp + i\,\frac{m}{M_1}\,D_{1T}^\perp \right),
\label{eom1}
\\[3mm]
\Delta_{A\,\alpha}^{\ [i\sigma^{\alpha -}\gamma_5 ]}
&=&  \frac{M_1}{z} \left( \tilde{H}_{s}
+ i\,\tilde{E}_s\right),
\\[3mm]
\Delta_A^{\alpha [\gamma^{-}]} + i\epsilon_T^{\alpha \beta}\,
\Delta_{A\,\beta}^{\ [\gamma^{-}\gamma_5]}
&=&
k_T^\alpha \left(\frac{1}{z}\tilde{D}^\perp +i\,\frac{m}{M_1}\,H_1^\perp
\right)
-\frac{\left(k_T^\alpha k_T^i + \frac{1}{2} k_T^2 g_T^{\alpha i}\right)}{M_1}
\,\epsilon_T^{ij}S_{1T\,j}^{}\,D_{1T}^\perp
\nonumber \\[3mm]
&+& i\epsilon_T^{\alpha \beta}  k_{T \beta}^{} \,\frac{1}{z}\,
\left(\tilde{G}_{s}^\perp
-i\,\lambda_1\tilde{D}_L^\perp\right)
+ i\epsilon_T^{\alpha \beta}  S_{1T\,\beta}^{}\, \frac{M_1}{z} \left(
\tilde{G}_T^\prime -i\, \tilde{D}_T \right),
\label{eom3}
\end{eqnarray}
where the functions indicated with a tilde ($\tilde H$, $\tilde E$, \ldots)
differ from the corresponding twist-3 functions ($H$, $E$, \ldots) by a
twist-2 part, namely
\begin{eqnarray}
& & E=\frac{m}{M_1}z D_1 + \tilde{E},\\
& & D^\perp=z D_1 + \tilde{D}^\perp,\\
& & D_L^\perp=\tilde{D}_L^\perp,\\
& & D_T=-\frac{\bm{k}_T^2}{2 M_1^2}z D_{1T}^\perp + 
\tilde{D}_T,
\label{DTtildedef}\\
& & E_s=\tilde{E}_s,\\
& & G_T^\prime=\frac{m}{M_1}z H_{1T} +\tilde{G}_T^\prime,\\
& & G_s^\perp=z G_{1s} + \frac{m}{M_1}z H_{1s}^\perp +\tilde{G}_s^\perp,\\
& & G_T= \frac{\bm{k}_T^2}{2M_1^2}z G_{1T}
+\frac{m}{M_1}z H_1 +\tilde{G}_T,\\
& & H_T^\perp=z H_{1T}+ \tilde{H}_T^\perp,\\
& & H=-\frac{\bm{k}_T^2}{M_1^2}z H_1^\perp + \tilde{H},
\label{Htildedef}\\
& & H_s=\frac{m}{M_1} z G_{1s} - \frac{\bm{k}_T^{} 
\cdot \bm{S}_{1T}^{}}{M_1} z H_{1T}
-\frac{\bm{k}_T^2}{M_1^2} z H_{1s}^\perp + \tilde{H}_s.
\end{eqnarray}
We have included $G_T$ in this list since it is relevant for 
$\bm k^\prime_T$-integrated functions and note that $\tilde G_T$ = 
$\tilde G_T^\prime + (\bm{k}_T^2/2M_1^2)\,\tilde 
G_T^\perp$. The functions in Eqs.\ (\ref{eom1}) to (\ref{eom3}) are
interaction-dependent and vanish for the case of a 
quark fragmenting in a quark (as can be checked with the help of Appendix A). 
Note that the time-reversal odd
twist-2 functions\footnote{The arbitrariness in the definition of
$\tilde{D}_T$ and $\tilde{H}$ in Eqs.\ (\ref{DTtildedef}) and
(\ref{Htildedef}) is fixed by the requirement that the functions 
$D_{1T}^\perp$ and
$H_1^\perp$ do not appear in the integrated versions of Eqs.\ (\ref{eom1}) to
(\ref{eom3}).} $D_{1T}^\perp$ and $H_1^\perp$ are in fact 
interaction-dependent. Their presence is due
to final state interactions of the produced hadrons, which after all are
strong interactions. 
The separation of twist-3 functions in this way is analogous to the case of
the distribution function $g_T$ = $g_1 + g_2$, and the twist-2 parts could
be called Wandzura-Wilczek parts \cite{Wandzura-Wilczek-77}. 

For the fragmentation of an antiquark most things are analogous to the quark
fragmentation. The major difference in our case is that the role of the $+$ 
and $-$ direction is reversed. We will denote the antiquark correlation 
functions by
$\overline \Delta (P_2,S_2;p)$. These should be defined consistently
with the replacement $\psi \rightarrow \psi^c = C \overline{\psi}{}^T$, or
$\overline{\Delta}{}^{\, [\Gamma]}= \Delta^{c[\Gamma]}$ for $\Gamma=\gamma_\mu,
i\sigma_{\mu\nu} \gamma_5, i\gamma_5$ and $\overline{\Delta}{}^{\, [\Gamma]}=
-\Delta^{c[\Gamma]}$ for $\Gamma={\bf 1}, \gamma_\mu \gamma_5$, where we
have defined the projections as:
\begin{eqnarray}
\overline \Delta^{\, [\Gamma]}(\bar z,\bm{p}_T^{}) & = &
\left. \frac{1}{4\bar z}\int dp^-\ \text{Tr} (\overline \Delta\,\Gamma)
\right|_{p^+ = P_2^+/\bar z,\ \bm{p}_{\scriptscriptstyle T}},
\end{eqnarray}
where we make the following Sudakov decomposition for the antiquark momentum
$p$:
\begin{equation}
p \equiv \frac{z_{2} Q}{\bar z\sqrt{2}}\,n_+
+ \frac{\bar z (p^2 + \bm{p}_T^2)}{z_{2} Q 
\sqrt{2}}\,n_- + p_T^{}
 \approx  \frac{1}{\bar z} P_2 + p_T^{}.
\end{equation}
Similarly, we decompose the spin vector $S_2$:
\begin{equation}
S_2 \equiv \frac{\lambda_2 z_{2} Q}{M_2  \sqrt{2}}\,n_+
-\frac{\lambda_2 M_2}{z_{2} Q \sqrt{2}}\,n_- + S_{2T}^{}
 \approx  \frac{\lambda_2}{M_2} P_2 + S_{2T}^{},
\end{equation}
with for a pure state $\lambda_2^2+\bm{S}_{2T}^2=1$.
The antiquark fragmentation functions are denoted by 
$\overline D_1(\bar z,\bm{p}_T^\prime{}^2), \ldots$, 
with $\bm{p}_T^\prime=-\bar z
\bm{p}_T^{}$, in full 
analogy to
the quark fragmentation functions. The antiquark fragmentation functions are 
obtained from
\begin{equation}
\overline \Delta_{ij}(P_2,S_2;p) = \sum_X \frac{1}{(2\pi)^4}
\int d^4x\ e^{-ip\cdot x}\,
\langle 0 \vert \overline \psi_j(0) \vert P_2, S_2; X \rangle
\langle P_2,S_2; X \vert \psi_i(x) \vert 0 \rangle.
\end{equation}

Although the antiquark-gluon-antiquark correlation functions are 
straightforwardly defined, we still give here the relations which follow from 
the e.o.m., since these differ non-trivially from those for the 
quark-gluon-quark correlation functions. We will not use tilde functions here,
since in the non-symmetric frame in which we will express the hadron 
tensor (non-symmetric between
quark and antiquark fragmentation part), they do not show up in a natural way.
\begin{eqnarray}
\overline\Delta_{A\,\alpha}^{\ [\sigma^{\alpha +} ]} =
\epsilon_T^{\alpha \beta}\,\overline\Delta_A^{\,\alpha [i\sigma^{\beta +}
\gamma_5]}
&=& i\left(\frac{M_2}{\bar z}\overline E - m\,\overline D_1
+i\,\frac{M_2}{\bar z}\,\overline H
- i\,\frac{p_T^2} {M_2}\,\overline H_1^\perp\right)
\nonumber \\[3mm]
&+& \epsilon_T^{ij}\,p_{Ti}^{} S_{2T\,j}^{} \,
\left(\overline H_{1T} - \frac{1}{\bar z}\overline H_T^\perp
+i\,\frac{m}{M_2}\,\overline D_{1T}^\perp \right),
\\[3mm]
\overline \Delta_{A\,\alpha}^{\ [i\sigma^{\alpha +}\gamma_5 ]}
&=&  -\frac{M_2}{\bar z} \overline H_{s} + m\,\overline G_{1s}
+ i\,\frac{M_2}{\bar z}\overline E_s + (p_T^{} \cdot S_{2T}^{})\overline H_{1T}
+ \frac{p_T^2}{M_2}\,\overline H_{1s}^\perp ,
\\[3mm] \overline \Delta_A^{\, \alpha [\gamma^{+}]} - i\epsilon_T^{\alpha 
\beta}\,
\overline \Delta_{A\,\beta}^{\ [\gamma^{+}\gamma_5]}
&=& -p_T^\alpha \left(\frac{1}{\bar z}\overline D^\perp
- \overline D_1-i\,\frac{m}{M_2}\,\overline H_1^\perp\right)
-\frac{p_T^\alpha}{M_2}\,\epsilon_T^{ij}p_{Ti}^{} S_{2T\,j}^{}
\,\overline D_{1T}^\perp
\nonumber \\[3mm]
 &
- &i\epsilon_T^{\alpha \beta}p_{T\beta}^{}
\left( \frac{1}{\bar z}\overline G_{s}^\perp - \overline G_{1s}
- \frac{m}{M_2}\,\overline H_{1s}^\perp
+i\,\frac{\lambda_2}{\bar z}\overline D_L^\perp\right)
\nonumber \\[3mm] &
-& i\epsilon_T^{\alpha \beta}S_{2T\,\beta}^{}\left(\frac{M_2}{\bar z}
\,\overline G_T^\prime - m\, \overline H_{1T} +i\,\frac{M_2}{\bar z}
\,\overline D_T\right).
\end{eqnarray}

Until now we have not commented on color gauge invariance of the correlation 
functions. As given above they are gauge-invariant quantities
displayed in a specific gauge. In general, one has to include 
path-ordered exponentials, 
in order to compensate for the gauge non-invariance due to
the non-locality of the operators. Such a link operator is of the form:
\begin{equation}
{\cal L}(0,x) = {\cal P} \exp \left(-ig\int_0^{x} dz^\mu \,A_\mu(z)\right).
\end{equation}
At this point we assume that matrix elements with multiple $A^-$-gluon
fields in $\Delta^{-}_{A}, \Delta^{--}_{AA}, \ldots$ (multiple 
$A^+$-gluon fields 
in $\overline \Delta^{\,+}_{\,A}, \overline \Delta^{\,++}_{\,AA}, \ldots$) 
will combine into an appropriate link operator with path along the $+$  
direction ($-$ direction) in $\Delta$
($\,\overline \Delta\,$) (cf.\ \cite{Qiu-Sterman-91}). 
For the $\bm{k}_T^{}$-dependent functions which involve 
transverse 
separations, the path from the point $0$ to $x$ in $\Delta$ will run along
the + direction via 
$x^+ = \infty$. The transverse part of the path, which is at $\infty$, 
does not contribute, since matrix elements are assumed to vanish there.

There remains 
one issue to be addressed, namely the explicit $A_T^{}$ in $\Delta_A^\alpha$ 
is not gauge invariant. Nevertheless, using the covariant 
derivative,
one can express this $A_T^{}$ in terms of $D_T^{}$ and $\partial_T^{}$, such 
that:
\begin{equation}
\Delta_{A}^{\alpha [\Gamma]} (z,\bm{k}_T^{}) = 
\Delta_{D}^{\alpha [\Gamma]} 
(z,\bm{k}_T^{}) - k^\alpha \,\Delta^{[\Gamma]} 
(z,\bm{k}_T^{}),
\end{equation}
where
\begin{equation}
\sum_X \frac{1}{(2\pi)^4}\int d^4x \ e^{ik\cdot x}\,
\langle 0 \vert \psi_i(x)\,i\partial^\mu\,\vert P_1,S_1; X \rangle
\langle P_1, S_1; X \vert\,\overline \psi_j(0)
\vert 0 \rangle = k^\mu \,\Delta_{ij}(P_1,S_1;k).
\end{equation}
Hence, we see that (again after inclusion of a link operator, 
assumed to arise from $\Delta_{AA}^{\alpha-}, \Delta_{AAA}^{\alpha--}, 
\ldots$) $\Delta_{A}^{\alpha}$ 
is a color gauge invariant quantity and therefore, so are the 
interaction-dependent parts of the twist-three fragmentation functions.

\section{The complete tree-level calculation}

\begin{figure}[htb]
\begin{center}
\leavevmode \epsfxsize=6cm \epsfbox{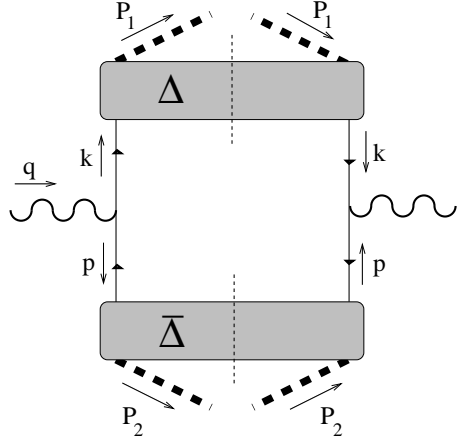}
\caption{\label{Leading} Quark diagram contributing to $e^+ e^-$ annihilation 
in leading order. There is a similar diagram with reversed fermion flow.}
\end{center}
\end{figure}

\begin{figure}[htb]
\begin{center}
\leavevmode \epsfxsize=10cm \epsfbox{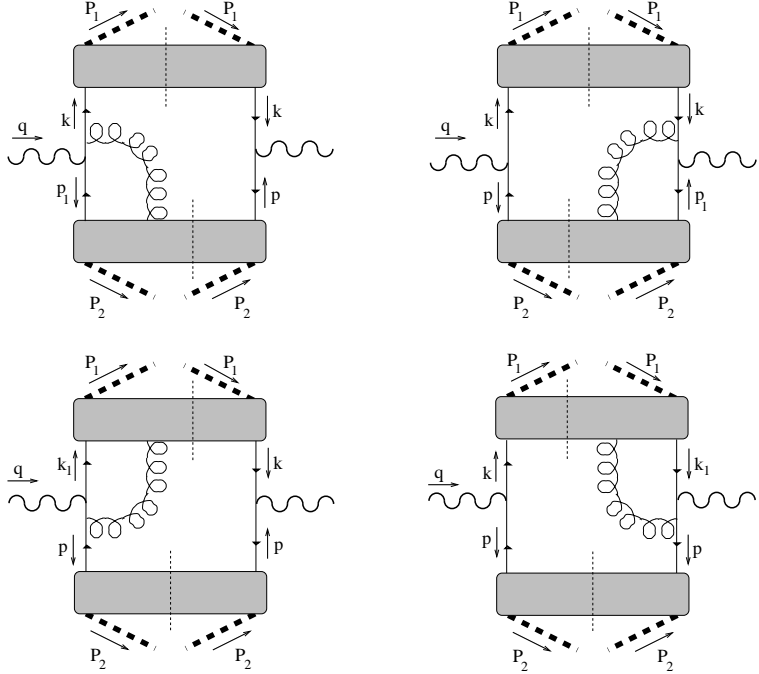}
\caption{\label{Subleading} Diagrams contributing to $e^+ e^-$ 
annihilation at order $1/Q$.}
\end{center}
\end{figure}

Up to order $1/Q$ there are five tree-level diagrams to consider. The simplest
diagram (Fig.\ \ref{Leading}) involving only quarks contributes at order $1$ 
and $1/Q$, 
the other four (Fig. \ref{Subleading}) involve one gluon which connects to one
of the two 
soft parts. 
Note that one power of the coupling constant is included in
the definition of the soft part, such that the diagrams are of order 
$(\alpha_s)^0$. 

The momentum conserving delta-function at the photon vertex is written 
as (neglecting $1/Q^2$ contributions)
\begin{equation} 
\delta^4(q-k-p)=\delta(q^+-p^+)\, \delta(q^--k^-)\, \delta^2(
\bm{p}_T^{}+
\bm{k}_T^{}-\bm{q}_T^{}),
\label{deltafn}
\end{equation}
fixing $P_2^+/\bar z=p^+=q^+=P_2^+/z_2$ and $P_1^-/z=k^-=q^-=P_1^-/z_1$. 
Eq.\ (\ref{deltafn}) shows why only the $k^+$ and 
$p^-$-integrated correlation functions are relevant. 
Note that the quark transverse
momentum integrations are linked.  
The five diagrams lead to the following expression for the full result up to 
order $1/Q$:
\begin{eqnarray}
{\cal W}^{\mu\nu}&=&3 e^2\int dp^- dk^+ d^2\bm{p}_T^{} d^2 
\bm{k}_T^{}\, \delta^2(\bm{p}_T^{}+
\bm{k}_T^{}-\bm{q}_T^{})\, \Biggl\{ 
\text{Tr}\left( \overline \Delta (p) 
\gamma^\mu \Delta 
(k) \gamma^\nu \right) \nonumber \\[3mm]
&-& \text{Tr}\left( \overline \Delta_A^\alpha (p) \gamma^\mu \Delta 
(k) \gamma_\alpha \frac{\slsh{n_+}}{Q\sqrt{2}} \gamma^\nu \right) 
- \text{Tr}\left( \left(\gamma_0 \overline \Delta_A^{\alpha\dagger} (p) 
\gamma_0
\right) \gamma^\mu
 \frac{\slsh{n_+}}{Q\sqrt{2}} \gamma_\alpha \Delta (k) \gamma^\nu \right)
\nonumber \\[3mm]
&+& \left. \text{Tr} \left( \overline \Delta (p) \gamma^\mu \left(\gamma_0
 \Delta_A^{\alpha\dagger} (k) \gamma_0\right) \gamma^\nu 
\frac{\slsh{n_-}}{Q\sqrt{2}} \gamma_\alpha 
\right) 
+ \text{Tr}\left( \overline \Delta(p) \gamma_\alpha
 \frac{\slsh{n_-}}{Q\sqrt{2}} \gamma^\mu \Delta_A^\alpha (k) \gamma^\nu 
\right) \Biggr\} \right|_{p^+\,k^-}. 
\label{Wmunustart}
\end{eqnarray}
The factor 3 originates from the color summation.
We have omitted the flavor indices and summation; furthermore, there is a 
contribution from diagrams with reversed fermion flow, which results from the 
above expression by replacing $\mu \leftrightarrow \nu$ and $q \rightarrow
-q$.  

In the expression the terms with $\slsh{n_\pm}$ arise from the fermion 
propagators in the hard part neglecting contributions that will appear 
suppressed by powers of $Q^2$, 
\begin{eqnarray}
\frac{\slsh{q}-\slsh{p_1}+m}{(q-p_1)^2-m^2} &\approx& 
\frac{(q^+-p_1{}^+)\gamma^-}{2(q^+-p_1{}^+)q^-}=\frac{\gamma^-}{2q^-}
=\frac{\slsh{n_+}}{Q\sqrt{2}},
\nonumber \\[3mm]
\frac{\slsh{k_1}-\slsh{q}+m}{(k_1-q)^2-m^2} &\approx& 
\frac{(k_1{}^--q^-)\gamma^+}{-2(k_1{}^--q^-)q^+}=\frac{\gamma^+}{-2q^+}
=-\frac{\slsh{n_-}}{Q\sqrt{2}},
\end{eqnarray}
where the approximate sign holds true only when the propagators are embedded 
in the diagrams. 
The quantity $\Delta_A^\alpha (k)$ arises from integrating out the second 
argument of $\Delta_A^\alpha (k,k_1)$ instead of the first which yields
the combination $\gamma_0 \Delta_A^{\alpha\dagger} (k) \gamma_0$:
\begin{eqnarray}
\int \frac{d^4k_1}{(2\pi)^4} \Delta_{A\, ij}^\alpha(P_1,S_1;k,k_1) &=&
\sum_X \frac{1}{(2\pi)^4}\int d^4x\
e^{i\,k\cdot x} \langle 0 \vert \psi_i(x)\, g A_T^\alpha (x)\,\vert P_1,S_1; X 
\rangle 
\langle P_1, S_1; X \vert 
\overline \psi_j(0) \vert 0 \rangle \nonumber \\[2mm]
&=& \Delta^\alpha_{A\, ij}(P_1,S_1;k), \\[2mm]
\int \frac{d^4k_1}{(2\pi)^4} \Delta_{A\, ij}^\alpha(P_1,S_1;k_1,k) &=&
\sum_X \frac{1}{(2\pi)^4} \int d^4x\ e^{i\,k\cdot x} \langle 0 \vert 
\psi_i(x)\,\vert P_1,S_1; X \rangle 
\langle P_1, S_1; X \vert \,g A_T^\alpha (0)\,
\overline \psi_j(0) \vert 0 \rangle \nonumber \\[2mm]
&=& (\gamma_0\Delta^{\alpha \dagger}_{A}\gamma_0)_{ij}(P_1,S_1;k)
\end{eqnarray} 
and similarly for $\overline \Delta_A^{\alpha} (p)$ and 
$\gamma_0 \overline \Delta_A^{\alpha\dagger} (p) \gamma_0$.
To deal with these combinations one can use the relation:
\begin{equation}
\left(\gamma_0\Delta^{\alpha \dagger}_{A}\gamma_0\right)^{[\Gamma]} = \left( 
\Delta^{\alpha [\Gamma]}_{A} \right)^*
\end{equation} 
and a similar one for $\gamma_0 \overline \Delta_A^{\alpha\dagger} (p) 
\gamma_0$.

To obtain the expressions for the symmetric and antisymmetric 
parts of the hadron tensor we expand all vectors in $\Delta, \overline \Delta,
\Delta_A^\alpha$ and 
$\overline{\Delta}{}_A^\alpha$ 
in the perpendicular basis ($\hat t$, $\hat z$ and
$\perp$ directions). In particular, we reexpress the transverse
vectors $k_T^{}$, $p_T^{}$, 
$S_{1T}^{}$ and $S_{2T}^{}$ in terms of 
their
perpendicular parts and a part along $\hat t$ and $\hat z$. For this we
need
\begin{equation}
g_T^{\mu\nu} = g_\perp^{\mu \rho}
g_{T\rho}^{\ \ \nu} 
- \frac{Q_T^{}}{Q}\,(\hat t^\mu + \hat z^\mu)\hat h^\nu.
\end{equation}
We will refer to the perpendicular projections as $k_\perp$, etc.\ (instead 
of the fully logical name, which would be $k_{T\perp}^{}$). 
Thus
\begin{equation}
k_\perp^\mu \equiv g_\perp^{\mu\nu}k_{T\nu}^{} = k_T^\mu 
+ \frac{\bm{q}_T^{}\cdot 
\bm{k}_T^{}}{Q}\,(\hat t^\mu + \hat z^\mu),
\end{equation}
and similarly for $p_\perp$, $S_{1\perp}$ and $S_{2\perp}$. We note that 
for these four vectors with this definition the two-component perpendicular 
parts are the same as the two-component transverse parts, i.e., 
$\bm k_\perp = \bm k_T^{}$, $\bm{S}_{1\perp} = \bm{S}_{1T}^{}$, etc. 

The full expressions for the symmetric and antisymmetric parts of the hadron 
tensor (expressed in the perpendicular frame defined in section 2) are 
given in Appendix B. We note that the expressions are not symmetric 
in the interchange of the hadrons $1$ and $2$, because the choice of
perpendicular direction ($P_{2\perp} \equiv 0$) is non-symmetric.

The cross-sections are obtained from the hadron tensor after contraction with
the lepton tensor
\begin{eqnarray}
L^{\mu \nu} & = &Q^2 \Biggl[
- \left( 1 - 2y + 2y^2 \right) g_\perp^{\mu \nu}
+ 4y(1-y) \hat z^\mu \hat z^\nu
\nonumber \\ && \qquad
-4y(1-y)\left( \hat \ell^\mu\hat \ell^\nu +\frac{1}{2}\,g_\perp^{\mu \nu}
\right)
- 2(1 - 2y)\sqrt{y(1-y)}\,\,\hat z_\ph^{\{ \mu}\hat \ell^{\nu \}}
\nonumber \\ &&\qquad
+i\lambda_e\,(1-2y)\,\epsilon_\perp^{\mu \nu}
- 2i\lambda_e\,\sqrt{y(1-y)}\,\,\hat \ell{}_\rho \epsilon_\perp^{\rho\,[ \mu}
\hat z_\ph^{\nu ]} \Biggr],
\end{eqnarray}
where $\{ \mu\nu \}$ indicates symmetrization of indices and $[ \mu\nu ]$ 
indicates antisymmetrization. The fraction $y$ is defined to be $y=
P_2 \cdot l/ P_2 \cdot q \approx l^-/q^-$, which in the lepton center of mass 
frame equals 
$y=(1 \pm \cos \theta_2)/2$, where 
$\theta_2$ is the angle of hadron 2 with respect to the momentum of the 
incoming leptons.
The contractions of specific tensor structures in the hadron tensor, given in 
Table 1,
contain azimuthal angles inside the perpendicular plane defined with respect 
to $\hat l_\perp^\mu$, defined to be the normalized perpendicular part of the 
lepton momentum $l$, $\hat l_\perp^\mu = l_\perp^\mu /(Q \sqrt{y(1-y)})$:
\begin{eqnarray}
\hat{l}_\perp \cdot a_\perp &=& - |{\bm a}_\perp | \cos \phi_a,
\label{anglecos} \\
\epsilon^{\mu\nu}_\perp \hat{l}_{\perp\mu} a_{\perp\nu}&=& |{\bm a}_\perp | 
\sin 
\phi_a.
\label{anglesin}
\end{eqnarray}

\begin{table}[htb]
\caption{\label{contractions}
Contractions of the lepton tensor $L_{\mu \nu}$ with tensor
structures appearing in the hadron tensor.}
\begin{tabular}{cl}
$w^{\mu \nu}$ & $L_{\mu \nu} w^{\mu \nu}/(4 Q^2)$ \\
& \\ \hline
$-g_\perp^{\mu \nu}$ &
$ \left( \frac{1}{2} - y + y^2 \right)$ \\
& \\   
$a_\perp^{\,\{ \mu} b_\perp^{\nu\}} -
(a_\perp \cdot b_\perp)\,g_\perp^{\mu\nu}$ &
$- y \left( 1 - y \right)
\vert \bm{a}_\perp \vert \, \vert \bm{b}_\perp \vert\,\cos(\phi_a+\phi_b)$ \\
& \\   
$\frac{1}{2} \left(
a_\perp^{\{\mu}\,\epsilon_\perp^{\nu \} \rho} b_{\perp\rho}
+ b_\perp^{\{\mu}\,\epsilon_\perp^{\nu \} \rho} a_{\perp\rho}\right)$ &
$ y\left( 1 - y \right)
\vert \bm{a}_\perp \vert \, \vert \bm{b}_\perp \vert\,\sin(\phi_a+\phi_b)$ \\
= $a_\perp^{\{\mu}\,\epsilon_\perp^{\nu\} \rho} b_{\perp\rho}
-(\epsilon_\perp^{\rho\sigma}a_{\perp \rho} b_{\perp\sigma})\,g_\perp^{\mu\nu}$
& \\ & \\
$\hat z^{\,\{\mu}a_\perp^{\nu\}}$ &
$-( 1 - 2y)\sqrt{y(1-y)}\,\vert \bm{a}_\perp \vert\,\cos \phi_a$ \\
& \\
$\hat z^{\,\{\mu }\,\epsilon_\perp^{\nu\} \rho}\,a_{\perp\rho}$ &
$( 1 - 2y)\sqrt{y(1-y)}\,\vert \bm{a}_\perp \vert\,\sin \phi_a$ \\
& \\
$i\,\epsilon_\perp^{\mu \nu}$ &
$-\lambda_e\,\,\left(\frac{1}{2}-y\right)$ \\
& \\
$i\,a_\perp^{\,[ \mu} b_\perp^{\nu ]}$ &
$-\lambda_e\,\,\left(\frac{1}{2}-y\right)\,
\vert \bm{a}_\perp \vert \, \vert \bm{b}_\perp \vert\,\sin (\phi_b - 
\phi_a)$ \\
& \\
$i\,\hat z^{\,[ \mu} a_\perp^{\nu ]}$ &
$\lambda_e\,\,\sqrt{y(1-y)}\,
\vert \bm{a}_\perp \vert \,\sin \phi_a$ \\
& \\
$i\,\hat z^{\,[\mu}\,\epsilon_\perp^{\nu\,] \rho}a_{\perp\rho}$ &
$\lambda_e\,\,\sqrt{y(1-y)}\,
\vert \bm{a}_\perp \vert \,\cos \phi_a$ \\
\end{tabular}
\end{table}
  
\section{Integration over transverse photon momentum}

In the next sections we will discuss explicit expressions for cross-sections.
Instead of giving the complete cross-section, which can be obtained from the 
hadron tensor (Appendix B), we treat a number of special cases. In this 
section we consider cross-sections integrated over all transverse momenta. 

After integration over the transverse momentum of the photon (or equivalently 
over the perpendicular momentum of hadron one $\bm{P}_{1\perp}=-z_{1} 
\bm{q}_T^{}$),
the integrations over $\bm{k}_T^{}$ and 
$\bm{p}_T^{}$ in the hadron tensor (Eqs.\ 
(\ref{WmunuS}) and (\ref{WmunuA})) can be performed leading to
\begin{eqnarray} 
\lefteqn{
\int d^2{\bm q_T^{}}\;{\cal W}^{\mu\nu}_S
=12 e^2 z_1z_2 \sum_{a,\bar a}{e_a}^2 }
\nonumber\\ &&
\times\Bigg\{
       - g_\perp^{\mu\nu}  \Bigg[
            D_1\overline D_1
          - \lambda_1\lambda_2 G_1\overline G_1
          \Bigg]
       - \left(S_{1\perp}^{\{\mu}S_{2\perp}^{\nu\}}
             +g_\perp^{\mu\nu}\,\bm S_{1\perp}\!\cdot \!\bm S_{2\perp}\,
\right)  
          \Bigg[ H_1\overline H_1 \Bigg]
\nonumber\\ &&
     {}- 2\frac{\hat z_\ph^{\{\mu}S_{1\perp}^{\nu\}}}{Q} \lambda_2 \Bigg[
            {M_1}\frac{\tilde G_T}{z_1}\overline G_1
          + {M_2}H_1\frac{\overline H_L}{z_2}
          \Bigg]
     {}+ 2\frac{\hat z_\ph^{\{\mu}S_{2\perp}^{\nu\}}}{Q} \lambda_1 \Bigg[
            {M_2}G_1\frac{\overline G_T}{z_2}
          + {M_1}\frac{\tilde H_L}{z_1}\overline H_1
          \Bigg]
\nonumber\\ &&
     {}+ 2\frac{\hat z_\ph^{\{\mu}\epsilon_\perp^{\nu\}\rho}
S_{1\perp}{}_\rho}{Q}  \Bigg[
            {M_1}\frac{\tilde D_T}{z_1}\overline D_1
          + {M_2}H_1\frac{\overline H}{z_2}\;
          \Bigg]
     {}+ 2\frac{\hat z_\ph^{\{\mu}\epsilon_\perp^{\nu\}\rho}
S_{2\perp}{}_\rho}{Q}  \Bigg[
            {M_2}D_1\frac{\overline D_T}{z_2}
          + {M_1}\frac{\tilde H}{z_1}\overline H_1
          \Bigg]
\Bigg\}
\end{eqnarray} 
and
\begin{eqnarray} 
\lefteqn{
\int d^2{\bm q_T^{}}\;{\cal W}^{\mu\nu}_A
=12 e^2 z_1z_2 \sum_{a,\bar a}{e_a}^2 }
\nonumber\\ &&
\times\Bigg\{
       i\epsilon_\perp^{\mu\nu}  \Bigg[
            \lambda_1 G_1\overline D_1
          - \lambda_2 D_1\overline G_1
          \Bigg]
\nonumber\\ &&
     {}+ 2i\frac{\hat z_\ph^{[\mu}S_{1\perp}^{\nu]}}{Q} \lambda_2 \Bigg[
            {M_1}\frac{\tilde D_T}{z_1}\overline G_1
          - {M_2}H_1\frac{\overline E_L}{z_2}
          \Bigg]
     {}+ 2i\frac{\hat z_\ph^{[\mu}S_{2\perp}^{\nu]}}{Q} \lambda_1 \Bigg[
          - {M_2}G_1\frac{\overline D_T}{z_2}
          + {M_1}\frac{\tilde E_L}{z_1}\overline H_1
          \Bigg]
\nonumber\\ &&
     {}+ 2i\frac{\hat z_\ph^{[\mu}\epsilon_\perp^{\nu]\rho}
S_{1\perp}{}_\rho}{Q}  \Bigg[
            {M_1}\frac{\tilde G_T}{z_1}\overline D_1
          + {M_2}H_1\frac{\overline E}{z_2}\;
          \Bigg]
     {}+ 2i\frac{\hat z_\ph^{[\mu}\epsilon_\perp^{\nu]\rho}
S_{2\perp}{}_\rho}{Q}  \Bigg[
            {M_2}D_1\frac{\overline G_T}{z_2}
          + {M_1}\frac{\tilde E}{z_1}\overline H_1
          \Bigg]
\Bigg\}.
\end{eqnarray}
We have now included the summation over flavor indices and $e_a$ is the quark
charge in units of $e$. 
The fragmentation functions are flavor dependent and only depend on the
longitudinal momentum fractions, e.g.\ $D_1\overline D_1$ 
= $D_1^a(z_1)\,\overline{D}{}_1^a(z_2)$.
The result is expressed in terms of the
fragmentation functions which survive the 
$\bm{k}_T$-integration of the Dirac
projections of the correlation functions (cf.\ Eqs.\ (\ref{projectionsL}) to
(\ref{projectionsSL})): 
$D_1, G_1=G_{1L}, H_1, E, E_L, G_T, H, H_L, D_T$ \cite{Jaffe-Ji-93,Ji-94}. 

Note that the tilde functions arise naturally
in the quark fragmentation region. The reason that this does not occur for the
antiquark fragmentation is due to the non-symmetric choice of frame. This
non-symmetric feature only shows up at subleading order. The leading order is
symmetric, since $\epsilon_\perp^{\mu\nu}$ acquires a minus
sign, due to the interchange of the vectors $n_+$ and $n_-$.

From the hadron tensors we easily arrive at the following expressions for the
cross-sections, where we separate the cross-sections into parts for
unpolarized (O) and polarized (L) leptons:
\begin{eqnarray} 
\frac{d\sigma^O(e^+e^-\to h_1h_2X)}{d\Omega dz_1 dz_2}&=&
\frac{3\alpha^2}{Q^2}\;\sum_{a,\bar a}{e_a}^2\;\Bigg\{ 
       A(y) \bigg(
            D_1\overline D_1
          - {\lambda_1}{\lambda_2}G_1\overline G_1
         \bigg)
\nonumber\\ &&
     {}+ B(y)\;|\bm S_{1T}^{}|\;|\bm 
S_{2T}^{}|\;\cos(\phi_{S_1}+\phi_{S_2})\;\bigg(
            H_1\overline H_1
         \bigg)  
\nonumber\\ &&
     {}+C(y)\,D(y)\;|\bm S_{1T}^{}|\;\sin(\phi_{S_1})
\;\bigg(
            \frac{2M_1}{Q}\frac{\tilde D_T}{z_1}\overline D_1
          + \frac{2M_2}{Q}H_1\frac{\overline H}{z_2}
         \bigg)
\nonumber\\ &&
     {}+C(y)\,D(y)\;|\bm S_{2T}^{}|\;\sin(\phi_{S_2})
\;\bigg(
            \frac{2M_2}{Q}D_1\frac{\overline D_T}{z_2}
          + \frac{2M_1}{Q}\frac{\tilde H}{z_1}\overline H_1
         \bigg)
\nonumber\\ &&
     {}+C(y)\,D(y)\;{\lambda_2}\;|\bm S_{1T}^{}|
\;\cos(\phi_{S_1})\;\bigg(
            \frac{2M_1}{Q}\frac{\tilde G_T}{z_1}\overline G_1
          + \frac{2M_2}{Q}H_1\frac{\overline H_L}{z_2}
         \bigg)
\nonumber\\ &&
     {}-C(y)\,D(y)\;{\lambda_1}\;|\bm S_{2T}^{}|\;
\cos(\phi_{S_2})\;\bigg(
            \frac{2M_2}{Q}G_1\frac{\overline G_T}{z_2}
          + \frac{2M_1}{Q}\frac{\tilde H_L}{z_1}\overline H_1
         \bigg)
\Bigg\}
\label{intxsO}
\end{eqnarray} 
and
\begin{eqnarray} 
\frac{d\sigma^L(e^+e^-\to h_1h_2X)}{d\Omega dz_1 dz_2}&=&
\frac{3\alpha^2}{Q^2}\;\lambda_e\;\sum_{a,\bar a}{e_a}^2\;\Bigg\{ 
         \frac{C(y)}{2}\bigg(
            {\lambda_2}D_1\overline G_1
          - {\lambda_1}G_1\overline D_1
         \bigg)
\nonumber\\ &&
     {}+ D(y)\;|\bm S_{2T}^{}|\;\cos(\phi_{S_2}) \;\bigg(
            \frac{2M_2}{Q}D_1\frac{\overline G_T}{z_2}
          + \frac{2M_1}{Q}\frac{\tilde E}{z_1}\overline H_1
         \bigg)
\nonumber\\ &&
     {}+ D(y)\;|\bm S_{1T}^{}|\;\cos(\phi_{S_1}) \;\bigg(
            \frac{2M_1}{Q}\frac{\tilde G_T}{z_1}\overline D_1
          + \frac{2M_2}{Q}H_1\frac{\overline E}{z_2}
         \bigg)
\nonumber\\ &&
     {}- D(y)\;{\lambda_1}\;|\bm S_{2T}^{}|\;
\sin(\phi_{S_2}) \;\bigg(
            \frac{2M_2}{Q}G_1\frac{\overline D_T}{z_2}
          - \frac{2M_1}{Q}\frac{\tilde E_L}{z_1}\overline H_1
         \bigg)
\nonumber\\ &&
     {}+ D(y)\;{\lambda_2}\;|\bm S_{1T}^{}|\;
\sin(\phi_{S_1}) \;\bigg(
            \frac{2M_1}{Q}\frac{\tilde D_T}{z_1}\overline G_1
          - \frac{2M_2}{Q}H_1\frac{\overline E_L}{z_2}
         \bigg)
\Bigg\},
\end{eqnarray}
where $d\Omega$ = $2dy\,d\phi^l$, with $\phi^l$ giving the 
orientation of $\hat l_\perp^\mu$, see Fig.~\ref{fig:kinann}. Note that 
on the r.h.s.\ of the above equations the dependence on $\phi^l$ enters in the 
azimuthal angles, which are defined with respect to $\hat l_\perp^\mu$, cf.\ 
Eqs.\ (\ref{anglecos}) and (\ref{anglesin}). We use the following factors:
\begin{eqnarray}
A(y) &=& \left(\frac{1}{2}-y+y^2\right),
\nonumber\\
B(y) &=& y\,(1-y),
\nonumber\\
C(y) &=& 1-2y,
\nonumber\\
D(y) &=& \sqrt{y\,(1-y)}.
\end{eqnarray}
The first three terms in Eq.\ (\ref{intxsO}) coincide with the ones found in
\cite{Chen-et-al-95}, if one neglects the contributions associated to Z
exchange in their Eq.\ (45). One observes that besides these three leading
contributions, one finds subleading single and double spin azimuthal
asymmetries.

To reduce the expression to the 1-particle inclusive cross-section, one
must take the fragmentation functions for a quark fragmenting into a quark
(see Appendix A) and sum over spins. Only $D_1^a(z_1)$ survives and after
summation over spins becomes
a delta-function. We find for the one-hadron
inclusive integrated cross-sections (using $h$ as running index instead 
of 2 and realizing that $\overline{D}{}_1^a = D_1^{\bar a}$):
\begin{eqnarray}
\frac{d\sigma^O(e^+e^-\to hX)}{d\Omega dz_h}&=&
\frac{3\alpha^2}{Q^2}\;\sum_{a,\bar a}{e_a}^2\;\Bigg\{
       A(y) D_1^a(z_h) +C(y)\,D(y) \;|\bm S_{hT}^{}|\;
\sin(\phi_{S_h})\;
            \frac{2M_h}{Q}\frac{D_T^a(z_h)}{z_h}
\Bigg\}
\label{int1PinclO}
\end{eqnarray}
and
\begin{eqnarray}
\frac{d\sigma^L(e^+e^-\to hX)}{d\Omega dz_h}&=&
\frac{3\alpha^2}{Q^2}\;\lambda_e\;\sum_{a,\bar a}{e_a}^2\;\Bigg\{
         - \frac{C(y)}{2} \; {\lambda_h} G_1^a(z_h)
+ D(y)\;|\bm S_{hT}^{}|\;\cos(\phi_{S_h}) \;
            \frac{2M_h}{Q}\frac{G_T^a(z_h)}{z_h}
       \Bigg\}.
\label{int1PinclL}
\end{eqnarray}
If the hadrons are unpolarized we find:
\begin{eqnarray}
\frac{d\sigma^O(e^+e^-\to hX)}{d\Omega dz_h}&=&
\frac{3\alpha^2}{Q^2}\,A(y)\;\sum_{a,\bar a}{e_a}^2\; D_1^a(z_h),
\\
\frac{d\sigma^O(e^+e^-\to h_1h_2X)}{d\Omega dz_1 dz_2}&=&
\frac{3\alpha^2}{Q^2}\,A(y)\;\sum_{a,\bar a}{e_a}^2\; D_1^a(z_1)
\overline {D}{}_1^a(z_2)
\end{eqnarray}
and $d\sigma^L=0$ in both cases. 
Hence we find for the number of produced particles 
\begin{eqnarray}
&&N_h(z_h)=\sum_{a,\bar a}{e_a}^2\; D_1^a (z_h) \Big/ \sum_{a,\bar a}{e_a}^2,\\
&&
N_{h_1 h_2}(z_1,z_2)=\sum_{a,\bar a}{e_a}^2\; D_1^a(z_1) 
\overline{D}{}_1^a(z_2) 
\Big/ \sum_{a,\bar a}{e_a}^2.
\end{eqnarray}
The case of $S_h$ = 0 gives the number of particles produced per spin degree 
of freedom, 
while the part proportional to $\lambda_h$ gives
the contributions of 
produced hadrons with $\lambda_h = \pm 1$. Thus the ratio of the
part multiplying $\lambda_h$ and the $S_h = 0$ result gives the
longitudinal polarization of the produced hadrons, which must lie 
between $-1$ and $+1$. Similarly, the ratio of
the part multiplying $\bm S_{hT}^{}$ and the $S_h = 0$ result
gives the transverse polarization, again a number lying between $-1$ and $+1$.
In many cases the final state hadron will not be a stable particle, e.g.\ a 
$\Lambda$. In that case the final state ($N\pi$ for the case of a $\Lambda$) 
is used to determine the spin vector $S_h$ \cite{Chen-et-al-95}.

For the 1-particle inclusive cross-section we see one leading polarizing 
effect, namely for polarized leptons the longitudinal polarization of 
produced spin-1/2 particles is given by
\begin{equation}
\Big( \mbox{longitudinal\ polarization} \Big) =
-\lambda_e\,\frac{C(y)}{2A(y)}\,\frac{\sum_{a,\bar a}{e_a}^2 
G_1^a(z_h)}{\sum_{a,\bar a}{e_a}^2 D_1^a(z_h)}.
\end{equation}
At subleading order transverse polarization in the final state is induced 
given by
\begin{eqnarray}
\left(\begin{array}{l} \mbox{transverse polarization} \\
\mbox{in\ lepton\ plane} \end{array}\right) & = &
\lambda_e\,\frac{D(y)}{A(y)}\,\frac{2M_h}{z_h\,Q}
\,\frac{\sum_{a,\bar a}{e_a}^2 G_T^a(z_h)}{\sum_{a,\bar a}{e_a}^2 D_1^a(z_h)},
 \\
\left(\begin{array}{l} \mbox{transverse polarization} \\
\mbox{transverse\ to\ lepton\ plane} \end{array}\right) & = &
\frac{C(y)\,D(y)}{A(y)}\,\frac{2M_h}{z_h\,Q}
\,\frac{\sum_{a,\bar a}{e_a}^2 D_T^a(z_h)}{\sum_{a,\bar a}{e_a}^2 D_1^a(z_h)},
\label{DTasym}
\end{eqnarray}
where the lepton plane is spanned by $l$ and $P_2$. 
The in-plane polarization is proportional to the lepton polarization and
is determined by the fragmentation function $G_T$. This function is 
the equivalent of the distribution function $g_T$. An out-of-plane 
polarization is found for unpolarized leptons, determined by the time 
reversal odd fragmentation function $D_T$. The asymmetry (\ref{DTasym}) was 
first discussed by Lu \cite{Lu-95}.

For the 2-particle inclusive cross-section in which one hadron has spin 
1/2, e.g.\ $e^+e^- \rightarrow \Lambda \pi X$, a longitudinal $\Lambda$ 
polarization is induced,
\begin{equation}
\Big( \mbox{longitudinal\ polarization} \Big)=
-\lambda_e\,\frac{C(y)}{2A(y)}\,\frac{\sum_{a,\bar a}{e_a}^2 
G_1^{a\rightarrow \Lambda}(z_1)
\,\overline{D}{}_1^{a\rightarrow \pi}(z_2)}
{\sum_{a,\bar a}{e_a}^2 D_1^{a\rightarrow \Lambda}(z_1) 
\overline{D}{}_1^{a\rightarrow \pi}(z_2)},
\end{equation}
involving one polarized fragmentation function, namely 
$G_1^{a\rightarrow\Lambda}$. 
If both hadrons have spin-1/2 a correlation between the polarizations of 
the two hadrons exist. The correlated longitudinal polarization involves 
$- {\lambda_1}{\lambda_2}\,A(y)\,\sum_{a,\bar a}{e_a}^2 G_1\overline G_1$; 
The correlated transverse polarization involves
$B(y)\;|\bm S_{1T}^{}|\;|\bm S_{2T}^{}|
\;\cos(\phi_{S_1}+\phi_{S_2})\;\sum_{a,\bar a}{e_a}^2 H_1\overline H_1$
and provides a
possibility to measure the transverse spin fragmentation function 
$H_1(z)$, the equivalent of the transverse spin distribution function 
$h_1$ \cite{Chen-et-al-95}. 
For the 2-particle inclusive cross-section there are several 
single spin asymmetries in unpolarized and polarized scattering, which 
among others give rise to twist 3 fragmentation functions $G_T$, $D_T$, 
$H_L$, $H$, $E$ and $E_L$, in principle each with characteristic final 
state polarization, but suppressed by $M_h/Q$.

\newpage

\section{Leading order asymmetries}

Instead of integrating out the $\bm{q}_T^{}$-dependence, 
we will now focus on the 
fully differential cross-section, i.e., not 
integrated over transverse momentum ($\bm{P}_{1\perp}=-z_{1} 
\bm{q}_T^{}$). We will see that the transverse momentum dependent
cross-sections contain asymmetries, which would vanish upon integration.
Some of those asymmetries appear at leading order, to which we restrict in this
section. The subleading results can be obtained from the hadron tensor 
in Appendix B in a similar way. 

First we consider the expression for both hadrons unpolarized:
\begin{equation}
\frac{d\sigma^O (e^+e^-\to h_1h_2X)}{d\Omega dz_1 dz_2 d^2{\bm q_T^{}}}=
\frac{3\alpha^2}{Q^2}\;z_1^2z_2^2\;\Bigg\{ 
          A(y)\;{\cal F}\left[D_1\overline D_1\right]
+ B(y)\;\cos(2\phi_1)\;
             {\cal F}\left[\left(2\,\hat{\bm{h}}\!\cdot \!
\bm k_T^{}\,\,\hat{\bm{h}}\!\cdot \!
\bm p_T^{}\,
                    -\,\bm k_T^{}\!\cdot \!
\bm p_T^{}\,\right)
                    \frac{H_1^{\perp}\overline H_1^{\perp}}{M_1M_2}\right]
\Bigg\},
\label{LO-OOO}
\end{equation}
where we use the convolution notation
\begin{equation} 
{\cal F}\left[D\overline D\, \right]\equiv\sum_{a,\bar a}{e_a}^2\;
\int d^2\bm k_T^{}\; d^2\bm p_T^{}\;
\delta^2 (\bm p_T^{}+\bm k_T^{}-\bm 
q_T^{})  D^a(z_{1},z_{1}^2 \bm{k}_T^2) 
\overline D^a(z_{2},z_{2}^2 \bm{p}_T^2),
\end{equation}
and $d\sigma^L =0$ in this case.
The angle $\phi_1$ is the azimuthal angle 
of $\hat h$, see Fig.\ 1.
So we find that the number of produced
hadrons has an azimuthal dependence:
\begin{eqnarray} 
\lefteqn{N_{h_1 h_2}(z_1,z_2,\bm{q}_T^{},y)= z_1^2z_2^2 
\;\Bigg\{ 
          {\cal F}\left[D_1\overline D_1\right]} \nonumber \\
&&\qquad \mbox{} + \frac{B(y)}{A(y)}\;\cos(2\phi_1)\;
             {\cal F}\left[\left(2\,\hat{\bm{h}}\!\cdot \!
\bm k_T^{}\,\,\hat{\bm{h}}\!\cdot \!
\bm p_T^{}\,
                    -\,\bm k_T^{}\!\cdot \!
\bm p_T^{}\,\right)
                    \frac{H_1^{\perp}\overline H_1^{\perp}}{M_1M_2}
\right] \Bigg\}\bigg/ \sum_{a,\bar a}{e_a}^2.
\label{reducedOO}
\end{eqnarray}
This asymmetry (the second term) has no analogue in the Drell-Yan 
process or semi-inclusive lepton-hadron scattering, since it involves a 
product of two time-reversal-odd functions. This new asymmetry goes 
with the same function $H_1^\perp$ as appears in the Collins effect, 
multiplied with the similar time-reversal-odd function $\overline H_1^\perp$. 
We emphasize that this is a measurement in which 
no polarization of the produced hadrons is needed and the result is
not suppressed by a factor of $1/Q$. This in contrast to the $\cos(2\phi)$
asymmetry found by Berger \cite{Berger-80}, which does not arise from
time-reversal-odd functions. It is $1/Q^2$ suppressed
and also arises in other processes.  

Assuming for instance a Gaussian 
$\bm{k}_T^{}$-dependence of the functions, 
the convolutions can be 
evaluated. The number of produced hadrons would then be:
\begin{eqnarray}
\lefteqn{N_{h_1 h_2}(z_1,z_2,Q_T^{},y)= 
{\cal G}(Q_T^{}; R) \sum_{a,\bar a}{e_a}^2\;
\Bigg\{ D_1^a(z_1) 
\overline{D}{}_1^a(z_2)} \nonumber\\
&& \qquad \mbox{} - \frac{B(y)}{A(y)}\;\cos(2\phi_1)\;
             \frac{Q_T^{} R^4}{M_1M_2 R_1^2 R_2^2} H_1^{\perp a}(z_1) 
\overline{H}{}_1^{\perp a}(z_2)
\Bigg\} \bigg/ \sum_{a,\bar a}{e_a}^2,
\end{eqnarray}
where $R^2=R_1^2 R_2^2 /(R_1^2 + R_2^2)$ and $D_1(z_1,
\bm{k}_T^\prime{}^2) = D_1(z_1) R_1^2 
\exp(-R_1^2 \bm{k}_T^2) / \pi z_1^2 \equiv 
D_1(z_1) {\cal G}(|\bm{k}_T^{} |; R_1)/ z_1^2$, etc. 
For details see Ref.\ 
\cite{Mulders-Tangerman-96}. 
 
In case we consider the expression for hadron one polarized and hadron two 
unpolarized, we find the following additional terms:
\begin{eqnarray} 
\lefteqn{
\frac{d\sigma^O (e^+e^-\to h_1h_2X)}{d\Omega dz_1 dz_2 d^2{\bm q_T^{}}}=
\frac{3\alpha^2}{Q^2}\;z_1^2z_2^2\;\Bigg\{ \ldots 
+ B(y)\;{\lambda_1}\;\sin(2\phi_1)\;
             {\cal F}\left[\left(2\,\hat{\bm{h}}\!\cdot \!
\bm k_T^{}\,\,\hat{\bm{h}}\!\cdot \!
\bm p_T^{}\,
                    -\,\bm k_T^{}\!\cdot \!
\bm p_T^{}\,\right)
                    \frac{H_{1L}^{\perp}\overline H_1^{\perp}}{M_1M_2}\right]}
\nonumber\\ && 
       \qquad - A(y)\;|\bm S_{1T}^{}|\;
\sin(\phi_1-\phi_{S_1})\; 
             {\cal F}\left[\,\hat{\bm{h}}\!\cdot \!\bm k_T^{}\,
                    \frac{D_{1T}^{\perp}\overline D_1}{M_1}\right]
        + B(y)\;|\bm S_{1T}^{}|\;\sin(\phi_1+\phi_{S_1})\;
             {\cal F}\left[\,\hat{\bm{h}}\!\cdot \!\bm p_T^{}\,
                    \frac{H_1\overline H_1^{\perp}}{M_2}\right]
\nonumber\\ &&
        \qquad + B(y)\;|\bm S_{1T}^{}|\;
\sin(3\phi_1-\phi_{S_1})\;
             {\cal F}\left[\left(
              4\,\hat{\bm{h}}\!\cdot \!\bm p_T^{}\,(\!\,
\hat{\bm{h}}\!\cdot \!\bm k_T^{}\,\!)^2
              -2\,\hat{\bm{h}}\!\cdot \!\bm k_T^{}\,\,
\bm k_T^{}\!\cdot \!\bm p_T^{}\,
              -\,\hat{\bm{h}}\!\cdot \!\bm p_T^{}\, \,
\bm k_T^2\,
                    \right)
                 \frac{H_{1T}^{\perp}\overline H_1^{\perp}}{2M_1{}^2M_2}\right]
\label{reducedO}
\Bigg\},\\[3mm]
\lefteqn{
\frac{d\sigma^L(e^+e^-\to h_1h_2X)}{d\Omega dz_1 dz_2 d^2{\bm q_T^{}}}=
\frac{3\alpha^2}{Q^2}\;z_1^2z_2^2\;\Bigg\{ 
        -\lambda_e\;\frac{C(y)}{2}\;\lambda_1\;
{\cal F}\left[G_1\overline D_1\right]}  
\nonumber\\ &&
   \qquad      -\lambda_e\;\frac{C(y)}{2}\;|\bm S_{1T}^{}|
\;\cos(\phi_1-\phi_{S_1})\;
             {\cal F}\left[\,\hat{\bm{h}}\!\cdot \!\bm k_T^{}\,
\frac{G_{1T}\overline D_1}{M_1}\right]
\label{reducedL}
\Bigg\}.
\end{eqnarray}
Again there is a term which has no analogue in semi-inclusive lepton-hadron 
scattering, namely the term with $D_{1T}^\perp$, which can be 
seen by comparison
with the result obtained in Ref.\ \cite{Tangerman-Mulders-95b}. The term with 
$H_1 \overline H_1^\perp$ is the analogue of the 
single-(transverse)-spin Collins effect. Note that it appears together with 
other single-transverse-spin asymmetries. 

Conversely, one can consider hadron two polarized and hadron one
unpolarized, which may be simpler from the experimental point of view, because
one does not need to measure the transverse momentum and the transverse 
polarization of
the same hadron. We find similar expressions in which all single spin 
terms have, besides the obvious replacements, a sign change.    

The leading order double spin asymmetries can be found in Appendix C and are
useful in for instance the case of $e^+e^- \rightarrow \Lambda \overline 
\Lambda X$. 

We like to point out that the transverse momentum dependence of some of 
the functions can be directly probed in the situation where hadron two 
is taken to be a jet, which in this back-to-back jet situation is equivalent 
to analyzing the azimuthal structure of hadrons 
inside a jet. 
Only $\overline D_1$ remains and is a delta-function, so the convolutions
can be evaluated exactly. In that case Eqs.\ (\ref{LO-OOO}), (\ref{reducedO})
and (\ref{reducedL}) taken together yield:
\begin{eqnarray} 
\frac{d\sigma(e^+e^-\to h \,\, \text{jet} \,\, 
X)}{d\Omega dz_h d^2{\bm q_T^{}}}
&=&\frac{3\alpha^2}{Q^2}\; z_h^2\;\sum_{a,\bar a}{e_a}^2\;\Bigg\{ 
          A(y)\; \left[ D_1^a(z_h,z_h^2 Q_T^2)
+ |\bm S_{hT}^{}|\;\sin(\phi_h-\phi_{S_1})\; 
            \frac{Q_T^{}}{M_h} D_{1T}^{\perp a}(z_h,z_h^2 Q_T^2) \right]
\nonumber\\ &&
        -\lambda_e\;\frac{C(y)}{2}\; \left[ \lambda_h\;G_1^a(z_h,z_h^2 Q_T^2)  
        + |\bm S_{hT}^{}|\;\cos(\phi_h-\phi_{S_1})\;
            \frac{Q_T^{}}{M_h} G_{1T}^a(z_h,z_h^2 Q_T^2) \right]
\label{reduced}
\Bigg\}.
\end{eqnarray}
This result means that there is a transverse polarization transverse to
the hadron plane, which is proportional to the function $D_{1T}^{\perp}$, and
a transverse polarization in the hadron plane, proportional to $G_{1T}$.

So one sees that by measuring $\bm{q}_T^{}$ one can learn 
about the transverse 
momentum dependence of four functions in this particular case. There
are no chiral-odd functions in this result, because they must be accompanied
by a quark mass, which gives a result proportional to $m/Q$, so they are 
present in the subleading result.

\section{Weighted cross-sections}

The expressions in the previous section contain convolutions,
which are not the objects of interest, rather we want the (universal) 
fragmentation functions depending on $z$ and $\bm{k}_T^2$. At the 
end of the previous section we discussed a situation in which
the transverse momentum dependence of some of the functions could be extracted
from the analysis of one jet. 
Below we will outline a way to obtain instead of the 
full transverse momentum dependence, the 
$\bm{k}_T^2$-moments of the functions, defined as:
\begin{equation}
F^{(n)}(z_{1})= \int d^2 \bm{k}_T^\prime \, \left(
\frac{\bm{k}_T^2}{2 M_1^2}\right)^n 
F(z_{1},\bm{k}_T^\prime{}^2),
\end{equation}
for a generic fragmentation function $F$. 
The lowest moment is the familiar $\bm{k}_T^{}$-integrated 
fragmentation function. By constructing appropriately weighted cross-sections
the convolutions result in products of such 
$\bm{k}_T^2$-moments.
The same $\bm{k}_T^2$-moments for
instance show up in semi-inclusive lepton-hadron scattering, in that case
multiplied by $\bm{k}_T^2$-moments of distribution
functions \cite{Mulders-Tangerman-96}. 
In section 5 we have presented the hadron tensor and cross-section
integrated over transverse photon momentum. A number of structures averaged
out to zero, which are retained when the integration is weighted with an
appropriate number of factors of $\bm{q}_T^{}$. 

We find for the once-weighted cross-sections, where we again show only the 
leading results, for the case that hadron two is unpolarized:
\begin{eqnarray} 
\lefteqn{
\int d^2{\bm q_T^{}}\;\left(\,\bm q_T^{}\!\cdot\!\bm a\,\right)\;
\frac{d\sigma(e^+e^-\to h_1h_2X)}{d\Omega dz_1 dz_2 d^2{\bm q_T^{}}}=
\frac{3\alpha^2}{Q^2}\;\sum_{a,\bar a}{e_a}^2\;|\bm{a}|\;\Bigg\{ }
\nonumber\\ &&
      \qquad - A(y)\;|\bm S_{1T}^{}|\;
\sin(\phi_{S_1}\!-\!\phi_a) \;\bigg(
            {M_1}D_{1T}^{\perp(1)}\overline D_1 \bigg) 
       - B(y)\;|\bm S_{1T}^{}|\;\sin(\phi_a+\phi_{S_1}) 
\;\bigg(
            {M_2}H_1
         \overline H_1^{\perp(1)} \bigg)
\nonumber\\ && \qquad -\lambda_e \frac{C(y)}{2}\;|\bm S_{1T}^{}|
\;\cos(\phi_{S_1}\!-\!\phi_a) \;\bigg(
            {M_1}G_{1T}^{(1)}\overline D_1
         \bigg) \Bigg\}.
\end{eqnarray}

Constructing from this cross-section the weighted one-particle inclusive
cross-section, by replacing $\overline D_1$ by a delta-function, and
considering the specific case $\bm{a}=\bm{\hat l}_\perp$, one finds:
\begin{eqnarray} 
\lefteqn{
\int d^2{\bm q_T^{}}\;\left(\,\bm q_T^{}\!\cdot\!\bm{\hat l}_\perp\,\right)\;
\frac{d\sigma(e^+e^-\to h X)}{d\Omega dz_h d^2{\bm q_T^{}}}=
\frac{3\alpha^2}{Q^2}\;\sum_{a,\bar a}{e_a}^2\;\Bigg\{ }
\nonumber\\ &&
       \qquad - A(y)\;|\bm S_{hT}^{}|\;
\sin(\phi_{S_h}) \; {M_h}D_{1T}^{\perp(1)}
-\lambda_e \frac{C(y)}{2}\;|\bm S_{hT}^{}|
\;\cos(\phi_{S_h}) \; {M_h}G_{1T}^{(1)} \Bigg\}.
\end{eqnarray}
These $\bm{k}_T^2$-moments $D_{1T}^{\perp(1)}$ and
$G_{1T}^{(1)}$ are related to the twist three functions $D_T$ and $G_T$ 
via 
\begin{eqnarray} 
D_T(z)&=& z^3\frac{d}{dz}\left[\frac{D_{1T}^{\perp(1)}(z)}{z}\right], \\
G_T(z)&=& G_1(z)
 -z^3\frac{d}{dz}\left[\frac{G_{1T}^{(1)}(z)}{z}\right],
\end{eqnarray} 
respectively \cite{Mulders-Tangerman-96}. 
An experimental verification of these relations by
comparing the above cross-section to the one-particle inclusive results 
Eqs.\ (\ref{int1PinclO}) and (\ref{int1PinclL}) would be very interesting. 

The twice-weighted cross-section is:
\begin{eqnarray} 
\lefteqn{
\int d^2{\bm q_T^{}}\;
\left(\,\bm q_T^{}\!\cdot\!\bm a\,\right)\;\left(\,\bm q_T^{}\!\cdot\!\bm b\,
\right)\;
\frac{d\sigma(e^+e^-\to h_1h_2X)}{d\Omega dz_1 dz_2 d^2{\bm q_T^{}}}=
\frac{3\alpha^2}{Q^2}\;\sum_{a,\bar a}{e_a}^2\;|\bm{a}|\;|\bm{b}|\;\Bigg\{ }
\nonumber\\ &&
     \qquad {}- A(y)\;\cos(\phi_b\!-\!\phi_a) \;\bigg(
            {M_2}^2D_1\overline D_1^{(1)}
          + {M_1}^2D_1^{(1)}\overline D_1
         \bigg)
\nonumber\\ &&
     \qquad {}+ 2B(y)\;{M_1}{M_2}\;\bigg[
         \cos(\phi_b+\phi_a)\;\bigg(
            H_1^{\perp(1)}\overline H_1^{\perp(1)}
         \bigg)
        +\sin(\phi_b+\phi_a)\;\bigg(
            {\lambda_1}\;H_{1L}^{\perp(1)}\overline H_1^{\perp(1)}
         \bigg)\bigg]
\nonumber\\ &&
      \qquad {} + \lambda_e \frac{C(y)}{2}\;\cos(\phi_b\!-\!\phi_a) \;\bigg(
          + {\lambda_1}\;{M_2}^2G_1\overline D_1^{(1)}
          + {\lambda_1}\;{M_1}^2G_1^{(1)}\overline D_1
         \bigg)
\bigg\}.
\label{twice}
\end{eqnarray}
In particular, one can use $(\bm{q}_T^{} \cdot
\bm{\hat l}_\perp)^2$, so one puts $\bm{a}=\bm{b}=\bm{\hat l}_\perp$ in the
above equation ($\phi_a=\phi_b=0$), 
such that in case both hadrons are unpolarized
\begin{eqnarray} 
\lefteqn{
\int d^2{\bm q_T^{}}\;
\left(\bm{q}_T^{} \!\cdot \! \bm{\hat l}_\perp\right)^2\;
\frac{d\sigma(e^+e^-\to h_1h_2X)}{d\Omega dz_1 dz_2 d^2{\bm q_T^{}}}=
\frac{3\alpha^2}{Q^2}\;\sum_{a,\bar a}{e_a}^2\;\Bigg\{ }
\nonumber\\ &&
     \qquad {}- A(y)\;\bigg(
            {M_2}^2D_1\overline D_1^{(1)}
          + {M_1}^2D_1^{(1)}\overline D_1
         \bigg) + 2B(y)\;{M_1}{M_2}\;
         H_1^{\perp(1)}\overline H_1^{\perp(1)}
\bigg\}.
\label{twicered}
\end{eqnarray}
Going back to the result for $N_{h_1 h_2}$ in Eq.\ (\ref{reducedOO}), one sees
that 
weighting that result only with $\cos(2\phi_1)$ would produce a convolution of
$H_1^\perp$ and $\overline 
H_1^\perp$, while the result above shows that including appropriate factors of 
$|\bm{q}_T^{}|$ produces a product of 
$\bm{k}_T^2$-moments
of fragmentation functions, in this case $H_1^{\perp (1)} \overline 
H_1^{\perp (1)}$. The $\bm{k}_T^2$-moments can be used in other processes 
where they also occur. The above is an illustration of a general procedure. 

In Appendix D and E we give the integrated once and twice-weighted 
hadron tensors, respectively, in case both hadrons are polarized.
The twice-weighted result is only given to leading order. 

\newpage

\section{Summary and conclusions}

We have presented the complete tree-level result up to order $1/Q$ for
inclusive two-hadron production in electron-positron
annihilation. We consider the situation where the two hadrons 
belong to different, back-to-back jets. Polarization in the initial and final 
states is 
included for the case of spin-1/2 hadrons. In case of spinless 
hadrons one will focus on the ones that are produced most abundantly, 
like $\pi$'s and $K$'s, 
which also serve to study flavor dependence of fragmentation functions (see 
for instance \cite{Schwiening-97}). For the case of spin-1/2 hadrons 
$\Lambda$'s seem
most appropriate due to their self-analyzing decays (see for instance 
\cite{Chen-et-al-95,Aleph-96}). Hadrons with higher 
spin, like $\rho$'s, are not considered, because in that case a spin vector 
is not sufficient to describe the spin states.  

We have restricted ourselves to the case of one photon exchange, since we are
interested in power corrections which are, most likely, negligible in regions 
of $Q^2$ where the annihilation into $Z$ bosons becomes important, i.e., LEP
energies. In 
forthcoming work we will investigate the inclusion of $Z$'s in the 
leading part of our result taking into account transverse momentum. 

We have worked in a
diagrammatic approach based upon analogy to the one developed by Ellis,
Furmanski and Petronzio for DIS. 
In this approach soft parts of a scattering process 
are treated as hadronic matrix
elements of non-local operators. 
The soft parts occuring in the process under
consideration are given
by quark-quark and quark-gluon-quark correlation functions. The latter 
are necessary to achieve electromagnetic gauge
invariance and are related to the quark-quark ones by use of the equations of
motion. All soft non-perturbative physics is then parametrized by a set of 
fragmentation functions. The leading ones can be interpreted as quark decay
functions. 

We have done our calculations at tree-level under the assumption 
that collinear divergences could be absorbed in the fragmentation functions 
and that factorization consequently holds. 
Moreover, for the interpretation of our results one should
keep in mind that loop corrections, in general, will lead to finite order
$({\alpha_s})^n$-corrections effecting the magnitude of observables and may 
also lead to non-zero contributions to observables not present at leading
order (a well-known example is $W_L$ in DIS).

Our results include among others the following:
\begin{itemize}
\item 
We have considered the cross-sections integrated over transverse momenta of
the produced hadrons for both, polarized and unpolarized beams up to
subleading order. Our result contains a few terms, which have
been found previously in a leading order analysis for unpolarized 
$e^+e^-$ annihilation \cite{Chen-et-al-95}. 

\item
In particular, we have focussed on the information obtainable by observing
transverse momentum of one of the produced hadrons 
(defined either relative to a 
jet-axis or relative to the momentum of a hadron in the second jet). Although
cross-sections differential in transverse momentum are not easy measurable,
they are of particular interest, since they contain leading order asymmetries,
which would vanish upon integration.

We have found a number of new unpolarized, single and double spin 
asymmetries. Often they have no analogues in (semi-inclusive) 
DIS or the Drell-Yan process, since they
involve products -- or to be more specific, convolutions -- of two
time-reversal odd fragmentation functions. In particular, the $\cos(2\phi)$
dependence discussed in Sect.\ 6 is most likely measurable, since it is not
suppressed by powers of $1/Q$ and does not involve polarization, neither of
the beams nor of the final states.  

One-hadron inclusive measurements supplied with the additional determination
of the jet axis gives direct access to the transverse momentum dependence of
some of the fragmentation functions. 

\item
We have discussed how convolutions of fragmentation functions can be
converted into products of their $k_T^{2}$-moments. This is achieved by
appropriate weighting the integration over the transverse momentum
dependence, in the spirit of a Fourier analysis. This is another way of 
retaining asymmetries,
which would vanish upon (non-weighted) integration. 
\end{itemize} 

As a final note, it can be seen from our results 
which extra asymmetries will show up in, 
for instance, the Drell-Yan process if one allows for time-reversal odd 
distribution functions. 
Such a single spin asymmetry is discussed
in \cite{Hammon}.
However, the presence of time-reversal odd distribution 
functions would require some factorization breaking mechanism, like
the one discussed in \cite{Anselmino}.  

In conclusion, we emphasize the possibilities provided by measuring azimuthal
asymmetries and higher twist contributions in $e^+e^-$ annihilation in order
to learn more about the structure of hadrons. 

\acknowledgments 
We thank A. Brandenburg, X. Ji and O. Teryaev for useful discussions. 
This work is part of the research program of the foundation for
Fundamental Research of Matter (FOM) and the National Organization
for Scientific Research (NWO).

\appendix

\section{The fragmentation functions for a quark fragmenting into a quark}

We consider
the correlation function for the case of a quark (with momentum $k$) 
fragmenting into a quark (with momentum $p$ and spin $s$), given by
\begin{equation}
\delta_{ij}(p,s;k) = u_i(k,s) \overline u_j(k,s) \delta^4(k-p)
= \frac{1}{2} \bigl((\slsh{k} + m)(1 + \gamma_5 \slsh{s}) \bigr)_{ij} 
\delta^4(k-p),
\end{equation}
where the momentum and spin of the quark are parametrized as
\begin{eqnarray}
k & = & \left[ k^-,\ \frac{\bm{k}_T^2 + m^2}{2k^-},
\ \bm{k}_T^{} \right], \\
s & = & \left[ \frac{\lambda_q \,k^-}{m},\ -\frac{m\,\lambda_q}{2k^-}
+\frac{\bm{k}_T^{} \cdot \bm{s}_{qT}^{}}
{k^-} +\frac{\lambda_q \,\bm{k}_T^2}{2m\,k^-},
\ \bm{s}_{qT}^{} + \frac{\lambda_q}{m} 
\bm{k}_T^{} \right]
\end{eqnarray}
in terms of a quark lightcone helicity $\lambda_q$ and a quark lightcone
transverse polarization $\bm{s}_{qT}^{}$. The projections become for twist two
\begin{eqnarray}
&& \delta^{[\gamma^-]}(k)
= \frac{1}{2} \delta \left( z - 1 \right)\delta^2 
(\bm{k}_T^{} - \bm{p}_T^{}), \\
&& \delta^{[\gamma^-\gamma_5]}(k)
= \frac{1}{2} \lambda_q\,\delta \left( z - 1 \right)\delta^2 
(\bm{k}_T^{} - \bm{p}_T^{}),
\\
&&\delta^{[i\sigma^{i-}\gamma_5 ]}(k)
= \frac{1}{2} \bm{s}_{qT}^i\,\delta 
\left( z - 1 \right)\delta^2 (\bm{k}_T^{} 
- \bm{p}_T^{}),
\end{eqnarray}
where $z = p^-/k^-$. For twist three we get 
\begin{eqnarray} 
&&\delta^{[1]}(k) = \frac{m}{2k^-}\,\delta \left( z - 1 \right)\delta^2 
(\bm{k}_T^{}
- \bm{p}_T^{}), \\  
&&\delta^{[\gamma^i]}(k)
= \frac{\bm{k}_T^i}{2k^-}\,\delta 
\left( z - 1 \right)\delta^2 (\bm{k}_T^{} 
- \bm{p}_T^{}),\\ 
&& \delta^{[\gamma^i\gamma_5]}(k)
= \frac{(m\,\bm{s}_{qT}^i 
+ \lambda_q\,\bm{k}_T^i)}{2k^-}\,\delta 
\left( z - 1 \right)\delta^2 (\bm{k}_T^{} 
- \bm{p}_T^{}),\\
&& \delta^{[i\sigma^{ij}\gamma_5 ]}(k)
= \frac{\bm{s}_{qT}^i\bm{k}_T^j 
- \bm{k}_T^i \bm{s}_{qT}^j}{2k^-}\,
\delta \left( z - 1 \right)\delta^2 (\bm{k}_T^{} 
- \bm{p}_T^{}), \\
&& \delta^{[i\sigma^{-+}\gamma_5 ]}(k)
= \frac{m\,\lambda_q - \bm{k}_T^{} \cdot 
\bm{s}_{qT}^{}}{2k^-}\,
\delta \left( z - 1 \right)\delta^2 (\bm{k}_T^{} 
- \bm{p}_T^{}).
\end{eqnarray}

\section{The complete expression for the hadron tensor}

The full expressions for the symmetric and antisymmetric parts of the hadron 
tensor are (expressed in the perpendicular frame defined in section 2) 

\begin{eqnarray} 
\lefteqn{
{\cal W}^{\mu\nu}_S=12 e^2 z_1z_2
\int d^2\bm k_T^{}\; d^2\bm p_T^{}\;
\delta^2 (\bm p_T^{}+\bm k_T^{}
         -\bm q_T^{})\Bigg\{ }
\nonumber\\ &&
       - g_\perp^{\mu\nu} \bigg[
            D_1\overline D_1
          - G_{1s}\overline G_{1s}
          +\frac{\epsilon_\perp^{\rho\sigma}k_{\perp\rho}S_{1\perp\sigma}}
            {M_1} D_{1T}^{\perp}\overline D_1
          -\frac{\epsilon_\perp^{\rho\sigma}p_{\perp\rho}S_{2\perp\sigma}}
            {M_2} D_1\overline D_{1T}^{\perp}
\nonumber\\ && \hspace{46mm}
        {}- \frac{\,\bm k_\perp\!\cdot\!\bm p_\perp\,
                  \,\bm S_{1\perp}\!\cdot\!\bm S_{2\perp}\,
                 -\,\bm p_\perp\!\cdot\!\bm S_{1\perp}\,
                  \,\bm k_\perp\!\cdot\!\bm S_{2\perp}\,}{M_1M_2}
             D_{1T}^{\perp}\overline D_{1T}^{\perp}
         \bigg]
\nonumber\\ &&
       -\left(S_{1\perp}^{\{\mu}S_{2\perp}^{\nu\}}
             +g_\perp^{\mu\nu}\,\bm S_{1\perp}\!\cdot\!\bm S_{2\perp}\,\right)
            H_{1T}\overline H_{1T}
       -\frac{k_\perp^{\{\mu}p_\perp^{\nu\}}
              +g_\perp^{\mu\nu}\,\bm k_\perp\!\cdot\!\bm p_\perp\,}{M_1M_2} 
         \left(
            H_{1s}^{\perp}\overline H_{1s}^{\perp}
          + H_1^{\perp}\overline H_1^{\perp}
         \right)
\nonumber\\ &&
       -\frac{k_\perp^{\{\mu}S_{2\perp}^{\nu\}}
               +g_\perp^{\mu\nu}\,\bm k_\perp\!\cdot\!\bm S_{2\perp}\,}{M_1}
           H_{1s}^{\perp}\overline H_{1T}
       -\frac{p_\perp^{\{\mu}S_{1\perp}^{\nu\}}
               +g_\perp^{\mu\nu}\,\bm p_\perp\!\cdot\!\bm S_{1\perp}\,}{M_2}
           H_{1T}\overline H_{1s}^{\perp}
\nonumber\\ &&
       + \frac{k_\perp^{\{\mu}\epsilon_\perp^{\nu\}\rho}p_{\perp\rho}
          +p_\perp^{\{\mu}\epsilon_\perp^{\nu\}\rho}k_{\perp\rho}}{2M_1M_2} 
         \left(
            H_{1s}^{\perp}\overline H_1^{\perp}
          - H_1^{\perp}\overline H_{1s}^{\perp}
         \right)
\nonumber\\ &&
       - \frac{k_\perp^{\{\mu}\epsilon_\perp^{\nu\}\rho}S_{2\perp\rho}
              +S_{2\perp}^{\{\mu}\epsilon_\perp^{\nu\}\rho}k_{\perp\rho}}{2M_1}
            H_1^{\perp}\overline H_{1T}
       + \frac{p_\perp^{\{\mu}\epsilon_\perp^{\nu\}\rho}S_{1\perp\rho}
              +S_{1\perp}^{\{\mu}\epsilon_\perp^{\nu\}\rho}p_{\perp\rho}}{2M_2}
            H_{1T}\overline H_1^{\perp}
\nonumber\\ &&
\nonumber\\ &&
     {}+ 2\frac{\hat z_\ph^{\{\mu}k_\perp^{\nu\}}}{Q} \bigg[
          + \frac{\tilde D^{\perp}}{z_1}\overline D_1
          - \frac{\,\bm S_{1\perp}\!\cdot\!\bm S_{2\perp}\,}{M_1}{M_2}
                 D_{1T}^{\perp}\frac{\overline D_T}{z_2}
          - {\lambda_2}\frac{\,\bm p_\perp\!\cdot\!\bm S_{1\perp}\,}{M_1}
                 D_{1T}^{\perp}\frac{\overline D_L^{\perp}}{z_2}
\nonumber\\ && \hspace{22mm}
          + \frac{\,\bm k_\perp\!\cdot\!\bm p_\perp\,
                  \,\bm S_{1\perp}\!\cdot\!\bm S_{2\perp}\,
                 -\,\bm p_\perp\!\cdot\!\bm S_{1\perp}\,
                  \,\bm k_\perp\!\cdot\!\bm S_{2\perp}\,}{M_1M_2}
                 D_{1T}^{\perp}\overline D_{1T}^{\perp}
\nonumber\\ && \hspace{22mm}
          - \frac{\tilde G_s^{\perp}}{z_1}\overline G_{1s}
          + \frac{\,\bm p_\perp\!\cdot\!\bm S_{1\perp}\,}{M_2}
                 \frac{\tilde H_T^{\perp}}{z_1}\overline H_{1s}^{\perp}
          + \,\bm S_{1\perp}\!\cdot\!\bm S_{2\perp}\,
               \frac{\tilde H_T^{\perp}}{z_1}\overline H_{1T}
          - \frac{{M_2}}{M_1}\left(
               H_{1s}^{\perp}\frac{\overline H_s}{z_2}
              +H_1^{\perp}\frac{\overline H}{z_2}\right)
         \bigg]
\nonumber\\ &&
\nonumber\\ &&
     {}+ 2\frac{\hat z_\ph^{\{\mu}p_\perp^{\nu\}}}{Q} \bigg[
          - D_1\frac{\overline D^{\perp}}{z_2}
          - \frac{\,\bm k_\perp^2\,\,\bm S_{1\perp}\!\cdot\!\bm S_{2\perp}\,}
                 {2M_1M_2}
                D_{1T}^{\perp}\overline D_{1T}^{\perp}
          + {\lambda_1}\frac{\,\bm k_\perp\!\cdot\!\bm S_{2\perp}\,}{M_2}
                \frac{\tilde D_L^{\perp}}{z_1}\overline D_{1T}^{\perp}
          + \frac{{M_1}}{M_2}\,\bm S_{1\perp}\!\cdot\!\bm S_{2\perp}\,
                \frac{\tilde D_T}{z_1}\overline D_{1T}^{\perp}
\nonumber\\ && \hspace{22mm}
        {}+ G_{1s}\frac{\overline G_s^{\perp}}{z_2}
          - \frac{\,\bm k_\perp\!\cdot\!\bm S_{2\perp}\,}{M_1}
                H_{1s}^{\perp}\frac{\overline H_T^{\perp}}{z_2}
          - \,\bm S_{1\perp}\!\cdot\!\bm S_{2\perp}\, 
                H_{1T}\frac{\overline H_T^{\perp}}{z_2}
          + \frac{{M_1}}{M_2}\left(
            \frac{\tilde H}{z_1}\overline H_1^{\perp}
          + \frac{\tilde H_s}{z_1}\overline H_{1s}^{\perp}\right)
         \bigg]
\nonumber\\ &&
\nonumber\\ &&
     {}+ 2\frac{\hat z_\ph^{\{\mu}S_{1\perp}^{\nu\}}}{Q} \bigg[
          + {\lambda_2}\frac{\,\bm k_\perp\!\cdot\!\bm p_\perp\,}{M_1}
              D_{1T}^{\perp}\frac{\overline D_L^{\perp}}{z_2}
          + \frac{{M_2}}{M_1}\,\bm k_\perp\!\cdot\!\bm S_{2\perp}\,
              D_{1T}^{\perp}\frac{\overline D_T}{z_2}
          - {M_1}\frac{\tilde G_T^{\prime}}{z_1}\overline G_{1s}
\nonumber\\ && \hspace{22mm}
        {}- \frac{\,\bm k_\perp\!\cdot\!\bm p_\perp\,}{M_2}
              \frac{\tilde H_T^{\perp}}{z_1}\overline H_{1s}^{\perp}
          - {M_2}H_{1T}\frac{\overline H_s}{z_2}
          - \,\bm k_\perp\!\cdot\!\bm S_{2\perp}\,
               \frac{\tilde H_T^{\perp}}{z_1}\overline H_{1T}
         \bigg]
\nonumber\\ &&
\nonumber\\ &&
     {}+ 2\frac{\hat z_\ph^{\{\mu}S_{2\perp}^{\nu\}}}{Q} \bigg[
          - {\lambda_1}\frac{\,\bm k_\perp\!\cdot\!\bm p_\perp\,}{M_2}
             \frac{\tilde D_L^{\perp}}{z_1}\overline D_{1T}^{\perp}
          + \frac{\,\bm p_\perp\!\cdot\!\bm S_{1\perp}\,\,\bm k_\perp^2\,}
              {2M_1M_2} D_{1T}^{\perp} \overline D_{1T}^{\perp}
          - \frac{{M_1}}{M_2}\,\bm p_\perp\!\cdot\!\bm S_{1\perp}\,
         \frac{\tilde D_T}{z_1}\overline D_{1T}^{\perp}
\nonumber\\ && \hspace{22mm}
        {}+ {M_2}G_{1s}\frac{\overline G_T^{\prime}}{z_2}
          + \frac{\,\bm k_\perp\!\cdot\!\bm p_\perp\,}{M_1}
               H_{1s}^{\perp}\frac{\overline H_T^{\perp}}{z_2}
          + \,\bm p_\perp\!\cdot\!\bm S_{1\perp}\,
               H_{1T}\frac{\overline H_T^{\perp}}{z_2}
          + {M_1}\frac{\tilde H_s}{z_1}\overline H_{1T}
         \bigg]
\nonumber\\ &&
\nonumber\\ &&
     {}+ 2\frac{\hat z_\ph^{\{\mu}\epsilon_\perp^{\nu\}\rho}k_{\perp\rho}}{Q} 
          \bigg[
            {\lambda_1}\frac{\tilde D_L^{\perp}}{z_1}\overline D_1
          - \frac{\,\bm p_\perp\!\cdot\!\bm S_{1\perp}\,}{M_1}
             D_{1T}^{\perp}\frac{\overline D^{\perp}}{z_2}
          - \frac{\,\bm k_\perp\!\cdot\!\bm S_{1\perp}\,}{M_1}
             D_{1T}^{\perp} \overline D_1
\nonumber\\ && \hspace{30mm}
        {}+ \frac{{m}}{M_1}H_1^{\perp}\overline G_{1s}
          - \frac{\,\bm p_\perp\!\cdot\!\bm S_{1\perp}\,}{M_2}
               \frac{\tilde H_T^{\perp}}{z_1}\overline H_1^{\perp}
          + \frac{{M_2}}{M_1}\left(
                H_{1s}^{\perp}\frac{\overline H}{z_2}
               -H_1^{\perp}\frac{\overline H_s}{z_2}\right)
         \bigg]
\nonumber\\ &&
\nonumber\\ &&
     {}+ 2\frac{\hat z_\ph^{\{\mu}\epsilon_\perp^{\nu\}\rho}p_{\perp\rho}}{Q} 
          \bigg[
            {\lambda_2}D_1\frac{\overline D_L^{\perp}}{z_2}
          - \frac{\,\bm k_\perp\!\cdot\!\bm S_{2\perp}\,}{M_2}
         \frac{\tilde D^{\perp}}{z_1}\overline D_{1T}^{\perp}
        {}- \frac{\,\bm k_\perp\!\cdot\!\bm S_{2\perp}\,}{M_1}
               H_1^{\perp}\frac{\overline H_T^{\perp}}{z_2}
          + \frac{{M_1}}{M_2}\left(
               \frac{\tilde H}{z_1}\overline H_{1s}^{\perp}
              -\frac{\tilde H_s}{z_1} \overline H_1^{\perp}\right)
         \bigg]
\nonumber\\ &&
\nonumber\\ &&
     {}+ 2\frac{\hat z_\ph^{\{\mu}\epsilon_\perp^{\nu\}\rho}
                S_{1\perp\rho}}{Q} \bigg[
            {M_1}\frac{\tilde D_T}{z_1}\overline D_1
          + \frac{\,\bm k_\perp\!\cdot\!\bm p_\perp\,}{M_2}
              \frac{\tilde H_T^{\perp}}{z_1}\overline H_1^{\perp}
          + \frac{\,\bm k_\perp\!\cdot\!\bm p_\perp\,}{M_1}
              D_{1T}^{\perp}\frac{\overline D^{\perp}}{z_2}
          + \frac{\,\bm k_\perp^2\,}{2M_1}
              D_{1T}^{\perp}\overline D_1
          + {M_2}H_{1T}\frac{\overline H}{z_2}
         \bigg]
\nonumber\\ &&
\nonumber\\ &&
     {}+ 2\frac{\hat z_\ph^{\{\mu}\epsilon_\perp^{\nu\}\rho}
                S_{2\perp\rho}}{Q} \bigg[
            {M_2}D_1\frac{\overline D_T}{z_2}
          + {M_1}\frac{\tilde H}{z_1}\overline H_{1T}
          + \frac{\,\bm k_\perp\!\cdot\!\bm p_\perp\,}{M_2}
         \frac{\tilde D^{\perp}}{z_1}\overline D_{1T}^{\perp}
          + \frac{\,\bm k_\perp\!\cdot\!\bm p_\perp\,}{M_1}
              H_1^{\perp}\frac{\overline H_T^{\perp}}{z_2}
         \bigg]
\Bigg\}
\label{WmunuS}
\end{eqnarray} 
\newpage
and
\begin{eqnarray} 
\lefteqn{
{\cal W}^{\mu\nu}_A=12 e^2 z_1z_2
\int d^2\bm k_T^{}\; d^2\bm p_T^{}\;
\delta^2 (\bm p_T^{}+\bm k_T^{}
         -\bm q_T^{})\Bigg\{ }
\nonumber\\ &&
     {}+ i\epsilon_\perp^{\mu\nu} \bigg[
            G_{1s}\overline D_1
          - D_1\overline G_{1s}
         \bigg]
       - ip_\perp^{[\mu}S_{2\perp}^{\nu]} 
            \frac{1}{M_2}G_{1s}\overline D_{1T}^{\perp}
       - ik_\perp^{[\mu}S_{1\perp}^{\nu]}
            \frac{1}{M_1}D_{1T}^{\perp}\overline G_{1s}
\nonumber\\ &&
\nonumber\\ &&
     {}+ 2i\frac{\hat z_\ph^{[\mu}k_\perp^{\nu]}}{Q} \bigg[
          + {\lambda_1}\frac{\tilde D_L^{\perp}}{z_1}\overline G_{1s}
          - \frac{\,\bm k_\perp\!\cdot\!\bm S_{1\perp}\,}{M_1}
               D_{1T}^{\perp}\overline G_{1s}
          - \frac{\,\bm S_{1\perp}\!\cdot\!\bm S_{2\perp}\,}{M_1}\left(
               {M_2}D_{1T}^{\perp}\frac{\overline G_T^{\prime}}{z_2}
             - {m}D_{1T}^{\perp}\overline H_{1T}\right)
\nonumber\\ && \hspace{23mm}
        {}- \frac{\,\bm p_\perp\!\cdot\!\bm S_{1\perp}\,}{M_1}\left(
               D_{1T}^{\perp}\frac{\overline G_s^{\perp}}{z_2}
              -\frac{{m}}{M_2}D_{1T}^{\perp}\overline H_{1s}^{\perp}\right)
          + \frac{{m}}{M_1}H_1^{\perp}\overline D_1
          - \frac{{M_2}}{M_1}\left(
                 H_{1s}^{\perp}\frac{\overline E_s}{z_2}
                +H_1^{\perp}\frac{\overline E}{z_2}\right)
         \bigg]
\nonumber\\ &&
\nonumber\\ &&
     {}+ 2i\frac{\hat z_\ph^{[\mu}p_\perp^{\nu]}}{Q} \bigg[
          - {\lambda_2}G_{1s}\frac{\overline D_L^{\perp}}{z_2}
          + \frac{\,\bm k_\perp\!\cdot\!\bm S_{2\perp}\,}{M_2}
               \frac{\tilde G_s^{\perp}}{z_1}\overline D_{1T}^{\perp}
          + \frac{\,\bm S_{1\perp}\!\cdot\!\bm S_{2\perp}\,}{M_2}{M_1}
             \frac{\tilde G_T^{\prime}}{z_1}\overline D_{1T}^{\perp}
\nonumber\\ && \hspace{23mm}
        {}+ \frac{{M_1}}{M_2}\left(
              \frac{\tilde E_s}{z_1}\overline H_{1s}^{\perp}
             +\frac{\tilde E}{z_1}\overline H_1^{\perp}\right)
         \bigg]
\nonumber\\ &&
\nonumber\\ &&
     {}+ 2i\frac{\hat z_\ph^{[\mu}S_{1\perp}^{\nu]}}{Q} \bigg[
          + {M_1}\frac{\tilde D_T}{z_1}\overline G_{1s}
          + \frac{\,\bm k_\perp\!\cdot\!\bm S_{2\perp}\,}{M_1}\left(
                {M_2}D_{1T}^{\perp}\frac{\overline G_T^{\prime}}{z_2}
               -{m}D_{1T}^{\perp}\overline H_{1T}\right)
          + \frac{\,\bm k_\perp^2\,}{2M_1}
                D_{1T}^{\perp}\overline G_{1s}
\nonumber\\ && \hspace{24.5mm}
        {}+ \frac{\,\bm k_\perp\!\cdot\!\bm p_\perp\,}{M_1M_2}\left(
                {M_2}D_{1T}^{\perp}\frac{\overline G_s^{\perp}}{z_2}
               -{m}D_{1T}^{\perp}\overline H_{1s}^{\perp}\right)
          - {M_2}H_{1T}\frac{\overline E_s}{z_2}
         \bigg]
\nonumber\\ &&
\nonumber\\ &&
     {}+ 2i\frac{\hat z_\ph^{[\mu}S_{2\perp}^{\nu]}}{Q} \bigg[
          - {M_2}G_{1s}\frac{\overline D_T}{z_2}
          - \frac{\,\bm p_\perp\!\cdot\!\bm S_{1\perp}\,}{M_2}{M_1}
             \frac{\tilde G_T^{\prime}}{z_1}\overline D_{1T}^{\perp}
          - \frac{\,\bm k_\perp\!\cdot\!\bm p_\perp\,}{M_2}
             \frac{\tilde G_s^{\perp}}{z_1}\overline D_{1T}^{\perp}
          + {M_1}\frac{\tilde E_s}{z_1}\overline H_{1T}
         \bigg]
\nonumber\\ &&
\nonumber\\ &&
     {}+ 2i\frac{\hat z_\ph^{[\mu}\epsilon_\perp^{\nu]\rho}k_{\perp\rho}}{Q} 
          \bigg[
            \frac{\tilde G_s^{\perp}}{z_1}\overline D_1
          - \frac{\tilde D^{\perp}}{z_1}\overline G_{1s}
          - \frac{\,\bm p_\perp\!\cdot\!\bm S_{1\perp}\,}{M_1M_2}{m}
                D_{1T}^{\perp}\overline H_1^{\perp}
          + \frac{{M_2}}{M_1}\left(
                H_{1s}^{\perp}\frac{\overline E}{z_2}
               -H_1^{\perp}\frac{\overline E_s}{z_2}\right)
         \bigg]
\nonumber\\ &&
\nonumber\\ &&
     {}+ 2i\frac{\hat z_\ph^{[\mu}\epsilon_\perp^{\nu]\rho}p_{\perp\rho}}{Q} 
          \bigg[
            D_1\frac{\overline G_s^{\perp}}{z_2}
          - G_{1s}\frac{\overline D^{\perp}}{z_2}
          - \frac{\,\bm k_\perp\!\cdot\!\bm S_{2\perp}\,}{M_1M_2}{m}
                H_1^{\perp}\overline D_{1T}^{\perp}
          + \frac{{M_1}}{M_2}\left(
                \frac{\tilde E}{z_1}\overline H_{1s}^{\perp}
               -\frac{\tilde E_s}{z_1}\overline H_1^{\perp}\right)
         \bigg]
\nonumber\\ &&
\nonumber\\ &&
     {}+ 2i\frac{\hat z_\ph^{[\mu}\epsilon_\perp^{\nu]\rho}S_{1\perp\rho}}{Q} 
          \bigg[
            {M_1}\frac{\tilde G_T^{\prime}}{z_1}\overline D_1
          + \frac{\,\bm k_\perp\!\cdot\!\bm p_\perp\,}{M_1M_2}{m}
                 D_{1T}^{\perp}\overline H_1^{\perp}
          + {M_2}H_{1T}\frac{\overline E}{z_2}
         \bigg]
\nonumber\\ &&
\nonumber\\ &&
     {}+ 2i\frac{\hat z_\ph^{[\mu}\epsilon_\perp^{\nu]\rho}S_{2\perp\rho}}{Q} 
           \bigg[
            {M_2}D_1\frac{\overline G_T^{\prime}}{z_2}
          + \frac{\,\bm k_\perp\!\cdot\!\bm p_\perp\,}{M_1M_2}{m}
                 H_1^{\perp}\overline D_{1T}^{\perp}
          + {M_1}\frac{\tilde E}{z_1}\overline H_{1T}
         \bigg]
\Bigg\}.
\label{WmunuA}
\end{eqnarray} 

\section{double spin asymmetries}

In this Appendix we give the azimuthal dependences of double spin
asymmetries, as can be observed, for instance, in $\Lambda\overline \Lambda$
production by determination of the polarizations of both observed hadrons. The
spin-independent and single-spin dependent parts of the cross-section are
given in Eqs.\ (\ref{LO-OOO}), (\ref{reducedO}) and (\ref{reducedL}).

\begin{eqnarray} 
\lefteqn{
\frac{d\sigma^{(2)}(e^+e^-\to h_1h_2X)}
     {d\Omega dz_1 dz_2 d^2{\bm q_T^{}}}=
\frac{3\alpha^2}{Q^2}\;z_1^2z_2^2\;\Bigg\{ 
       - \frac{A(y)}{2}\;\lambda_1\;\lambda_2\;
            {\cal F}\left[G_1 \overline G_1\right]}
\nonumber\\ &&
        -A(y)\;\lambda_1\;|\bm S_{2T}^{}|\;
          \cos(\phi_1-\phi_{S_2})\;
            {\cal F}\left[\,\bm{\hat h}\!\cdot \!\bm p_T^{}\,
             \frac{G_1 \overline G_{1T}}{M_2}\right]  
\nonumber\\ &&
        + \frac{A(y)}{2}\;|\bm S_{1T}^{}|\;
                          |\bm S_{2T}^{}|\;
                \cos(2\phi_1-\phi_{S_1}-\phi_{S_2})\;
             {\cal F}\left[\,\bm{\hat h}\!\cdot \!\bm k_T^{}\,
                    \,\bm{\hat h}\!\cdot \!\bm p_T^{}\,
                    \frac{D_{1T}^\perp \overline D_{1T}^\perp
                         -G_{1T} \overline G_{1T}}{M_1M_2}\right]
\nonumber\\ &&
        - \frac{A(y)}{2}\;|\bm S_{1T}^{}|\;
                          |\bm S_{2T}^{}|\;
                \cos(\phi_1-\phi_{S_1})\;\cos(\phi_1-\phi_{S_2})\;
             {\cal F}\left[\,\bm k_T^{}\!\cdot \!
                      \bm p_T^{}\,
                    \frac{D_{1T}^\perp \overline D_{1T}^\perp}{M_1M_2}\right]
\nonumber\\ &&
        - \frac{A(y)}{2}\;|\bm S_{1T}^{}|\;
                          |\bm S_{2T}^{}|\;
                \sin(\phi_1-\phi_{S_1})\;\sin(\phi_1-\phi_{S_2})\;
             {\cal F}\left[\,\bm k_T^{}\!\cdot \!
                      \bm p_T^{}\,
                    \frac{G_{1T} \overline G_{1T}}{M_1M_2}\right]
\nonumber\\ &&
        + \frac{B(y)}{2}\;|\bm S_{1T}^{}|\;
                          |\bm S_{2T}^{}|\;
                \cos(\phi_{S_1}+\phi_{S_2})\;
            {\cal F}\left[H_1 \overline H_1\right]
\nonumber\\ &&
        + B(y)\;\lambda_1\;|\bm S_{2T}^{}|\;
                \cos(\phi_1+\phi_{S_2})\;
            {\cal F}\left[\,\bm{\hat h}\!\cdot \!\bm k_T^{}\,
                   \frac{H_{1L}^\perp \overline H_1}{M_1}\right]
\nonumber\\ &&
        + \frac{B(y)}{2}\;\lambda_1\;\lambda_2\;\cos(2\phi_1)\;
            {\cal F}\left[\left(2\,\bm{\hat h}\!\cdot \!\bm k_T^{}\,
                          \,\bm{\hat h}\!\cdot \!\bm p_T^{}\,
                         -\,\bm k_T^{}\!\cdot \!
                            \bm p_T^{}\,\right)
                   \frac{H_{1L}^\perp \overline H_{1L}^\perp}{M_1M_2}\right]
\nonumber\\ &&
       + \frac{B(y)}{2}\;|\bm S_{1T}^{}|\;
                         |\bm S_{2T}^{}|\;
               \cos(2\phi_1-\phi_{S_1}+\phi_{S_2})\;
            {\cal F}\left[\left(2(\,\bm{\hat h}\!\cdot \!
                             \bm k_T^{}\,)^2
                          -\,\bm k_T^2\,\right)
                   \frac{H_{1T}^\perp \overline H_1}{M_1}\right]
\nonumber\\ &&
        + \frac{B(y)}{2}\;\lambda_2\;|\bm S_{1T}^{}|\;
                \cos(3\phi_1-\phi_{S_1})\;
            {\cal F}\left[\left(4\,\bm{\hat h}\!\cdot \!\bm p_T^{}\,
                          (\bm{\hat h}\!\cdot \!\bm k_T^{})^2
                        -2\,\bm{\hat h}\!\cdot \!\bm k_T^{}\,
                          \,\bm k_T^{}\!\cdot \!
                            \bm p_T^{}\,
                        -\,\bm{\hat h}\!\cdot \!\bm p_T^{}\,
                         \,\bm k_T^2\,\right)
              \frac{H_{1T}^\perp \overline H_{1L}^\perp}{M_1^2M_2}\right]
\nonumber\\ &&
        + \frac{B(y)}{8}\;|\bm S_{1T}^{}|\;
                          |\bm S_{2T}^{}|\;
                \cos(4\phi_1-\phi_{S_1}-\phi_{S_2})\;
            {\cal F}\Bigg[\Bigg(
               8(\bm{\hat h}\!\cdot \!\bm k_T^{})^2
                (\bm{\hat h}\!\cdot \!\bm p_T^{})^2
              -4\,\bm k_T^{}\!\cdot \!
                  \bm p_T^{}\,
                \,\bm{\hat h}\!\cdot \!\bm k_T^{}\,
                \,\bm{\hat h}\!\cdot \!\bm p_T^{}\,
\nonumber\\ && \hspace{34mm}
              -2(\bm{\hat h}\!\cdot \!\bm k_T^{})^2\,
                 \bm p_T^2\,
              -2(\bm{\hat h}\!\cdot \!\bm p_T^{})^2\,
                 \bm k_T^2\,
              +(\bm{\hat h}\!\cdot \!\bm k_T^{})^2
               (\bm{\hat h}\!\cdot \!\bm p_T^{})^2\Bigg)
             \frac{H_{1T}^\perp \overline H_{1T}^\perp}{M_1^2M_2^2}\Bigg]
\nonumber\\ &&
     -\lambda_e\;\frac{C(y)}{2}\;\lambda_1\;|\bm S_{2T}^{}|\;
                 \sin(\phi_1-\phi_{S_2})\;
          {\cal F}\left[\,\bm{\hat h}\!\cdot \!\bm p_T^{}\,
                 \frac{G_1 \overline D_{1T}^\perp}{M_2}\right]
\nonumber\\ &&
     -\lambda_e\;\frac{C(y)}{4}\;|\bm S_{1T}^{}|\;
                                 |\bm S_{2T}^{}|\;
          \sin(2\phi_1-\phi_{S_1}-\phi_{S_2})\;
          {\cal F}\left[\,\bm{\hat h}\!\cdot \!\bm k_T^{}\,
                 \,\bm{\hat h}\!\cdot \!\bm p_T^{}\,
                 \frac{D_{1T}^\perp \overline G_{1T}
                       +G_{1T} \overline D_{1T}^\perp}{M_1M_2}\right]
\nonumber\\ &&
     -\lambda_e\;\frac{C(y)}{2}\;|\bm S_{1T}^{}|\;
                                 |\bm S_{2T}^{}|\;
          \sin(\phi_1-\phi_{S_2})\cos(\phi_1-\phi_{S_1})\;
          {\cal F}\left[\,\bm k_T^{}\!\cdot \!
                   \bm p_T^{}\,
                 \frac{D_{1T}^\perp \overline G_{1T}}{M_1M_2}\right]
      +\quad {\;1\longleftrightarrow 2\;\choose 
              \;\bm{p}\longleftrightarrow\bm{k}\;} \quad
\Bigg\}
\end{eqnarray} 

\section{Integrated once-weighted hadron tensor}

We display the hadron tensor weighted with the factor 
$\left(\,\bm q_T^{}\!\cdot\!\bm a\,\right)$ and integrated
over the transverse photon momentum. The vector $\bm a$ is an arbitrary vector
like, for instance, $\hat{\bm l}_\perp$. 

\begin{eqnarray} 
\lefteqn{
\int d^2{\bm q_T^{}}\;
\left(\,\bm q_T^{}\!\cdot\!\bm a\,\right)\;
{\cal W}^{\mu\nu}_S=12 e^2 z_1z_2\times\Bigg\{ }
\nonumber\\ &&
       - g_\perp^{\mu\nu}  \Bigg[
          - \lambda_1\,\bm a\!\cdot\!\bm S_{2\perp}\, {M_2} 
               G_1\overline G_{1T}^{(1)}
          - \lambda_2\,\bm a\!\cdot\!\bm S_{1\perp}\, {M_1} 
               G_{1T}^{(1)}\overline G_1
\nonumber\\ && \hspace{12mm}
          + \epsilon_\perp^{\rho\sigma}a_\rho S_{1\perp\sigma}
            {M_1}D_{1T}^{\perp(1)}\overline D_1
          - \epsilon_\perp^{\rho\sigma}a_\rho S_{2\perp\sigma}
            {M_2}D_1\overline D_{1T}^{\perp(1)}
          \Bigg]
\nonumber\\ &&
     {}- \left(S_{1\perp}^{\{\mu}a_\ph^{\nu\}}
          + g_\perp^{\mu\nu}\,\bm a\!\cdot\!\bm S_{1\perp}\,\right)
            \lambda_2{M_2}H_1\overline H_{1L}^{\perp(1)}
       - \left(S_{2\perp}^{\{\mu}a_\ph^{\nu\}}
          + g_\perp^{\mu\nu}\,\bm a\!\cdot\!\bm S_{2\perp}\,\right)
            \lambda_1{M_1}H_{1L}^{\perp(1)}\overline H_1
\nonumber\\ &&
     {}+ \left(a_\ph^{\{\mu}\epsilon_\perp^{\nu\}\rho}S_{1\perp\rho}  
             + S_{1\perp}^{\{\mu}\epsilon_\perp^{\nu\}\rho}a_\rho \right)
           \frac{M_2}{2}H_1\overline H_1^{\perp(1)}
       - \left(a_\ph^{\{\mu}\epsilon_\perp^{\nu\}\rho}S_{2\perp\rho}
             + S_{2\perp}^{\{\mu}\epsilon_\perp^{\nu\}\rho}a_\rho\right)
          \frac{M_1}{2}H_1^{\perp(1)}\overline H_1
\nonumber\\ &&
     {}+ 2\frac{\hat z_\ph^{\{\mu}S_{1\perp}^{\nu\}}}{Q}
            \,\bm a\!\cdot\!\bm S_{2\perp}\, \Bigg[
          + {M_1}{M_2} D_{1T}^{\perp(1)}\frac{\overline D_T}{z_2}
          - {M_1}{M_2} \frac{\tilde G_T}{z_1}\overline G_{1T}^{(1)}
          - {M_2}^2 H_1\frac{\overline H_T^{(1)}}{z_2}
          - {M_1}^2 \frac{\tilde H_T^{\perp(1)}}{z_1}\overline H_1
          \Bigg]
\nonumber\\ &&
     {}+ 2\frac{\hat z_\ph^{\{\mu}S_{2\perp}^{\nu\}}}{Q}
            \,\bm a\!\cdot\!\bm S_{1\perp}\,\Bigg[
          - {M_1}{M_2} \frac{\tilde D_T}{z_1}\overline D_{1T}^{\perp(1)}
          + {M_1}{M_2} G_{1T}^{(1)}\frac{\overline G_T}{z_2}
          + {M_2}^2 H_1\frac{\overline H_T^{\perp(1)}}{z_2}
          + {M_1}^2 \frac{\tilde H_T^{(1)}}{z_1}\overline H_1
          \Bigg]
\nonumber\\ &&
     {}- 2\frac{\hat z_\ph^{\{\mu}a_\ph^{\nu\}}}{Q}  \Bigg[
            {M_2}^2D_1\frac{\overline D^{\perp(1)}}{z_2}
          - {M_1}^2\frac{\tilde D^{\perp(1)}}{z_1}\overline D_1
          - {M_1}{M_2} \,\bm S_{1\perp}\!\cdot\!\bm S_{2\perp}\,\left(
                 \frac{\tilde D_T}{z_1}\overline D_{1T}^{\perp(1)}
               - D_{1T}^{\perp(1)}\frac{\overline D_T}{z_2}
            \right)
\nonumber\\ && \hspace{20mm}
          + \lambda_1\lambda_2\left(
                  {M_1}^2\frac{\tilde G_L^{\perp(1)}}{z_1}\overline G_1
                - {M_2}^2G_1\frac{\overline G_L^{\perp(1)}}{z_2}
            \right)
          + \lambda_1\lambda_2{M_1}{M_2} \left(
                  H_{1L}^{\perp(1)}\frac{\overline H_L}{z_2}
                - \frac{\tilde H_L}{z_1}\overline H_{1L}^{\perp(1)}
            \right)
\nonumber\\ && \hspace{20mm}
        {}+ {M_1}{M_2} \left(
                  H_1^{\perp(1)}\frac{\overline H}{z_2} 
                - \frac{\tilde H}{z_1}\overline H_1^{\perp(1)}
            \right) 
          - \,\bm S_{1\perp}\!\cdot\!\bm S_{2\perp}\,\left(
                  {M_1}^2 \frac{\tilde H_T^\perp(1)}{z_1}\overline H_1
                - {M_2}^2 H_1\frac{\overline H_T^{\perp(1)}}{z_2}
            \right)
          \Bigg]
\nonumber\\ &&
     {}- 2\frac{\hat z_\ph^{\{\mu}\epsilon_\perp^{\nu\}\rho}a_\rho}{Q}  \Bigg[
          - \lambda_1{M_1}^2\frac{\tilde D_L^{\perp(1)}}{z_1}\overline D_1
          - \lambda_2{M_2}^2D_1\frac{\overline D_L^{\perp(1)}}{z_2}
          - \lambda_2{m}{M_1}H_1^{\perp(1)}\overline G_1
\\ && \hspace{24mm}
          + \lambda_1{M_1}{M_2} \left(
                   \frac{\tilde H_L}{z_1}\overline H_1^{\perp(1)}
                 - H_{1L}^{\perp(1)}\frac{\overline H}{z_2}
            \right)
          + \lambda_2{M_1}{M_2} \left(
                   H_1^{\perp(1)}\frac{\overline H_L}{z_2}
                 - \frac{\tilde H}{z_1}\overline H_{1L}^{\perp(1)}
            \right)
          \Bigg]
\Bigg\} \nonumber
\end{eqnarray} 
and
\begin{eqnarray} 
\lefteqn{
\int d^2{\bm q_T^{}}\;
\left(\,\bm q_T^{}\!\cdot\!\bm a\,\right)\;
{\cal W}^{\mu\nu}_A=12 e^2 z_1z_2\times\Bigg\{ }
\nonumber\\ &&
     {}+ i\epsilon_\perp^{\mu\nu}  \Bigg[
            {M_1}\,\bm a\!\cdot\!\bm S_{1\perp}\,G_{1T}^{(1)}\overline D_1
          - {M_2}\,\bm a\!\cdot\!\bm S_{2\perp}\,D_1\overline G_{1T}^{(1)}
          \Bigg]
\nonumber\\ &&
     {}+ iS_{1\perp}^{[\mu}a_\ph^{\nu]}  \Bigg[
            \lambda_2{M_1}D_{1T}^{\perp(1)}\overline G_1
          \Bigg]
       + iS_{2\perp}^{[\mu}a_\ph^{\nu]}  \Bigg[
           \lambda_1{M_2}G_1\overline D_{1T}^{\perp(1)}
          \Bigg]
\nonumber\\ &&
     {}+ i2\frac{\hat z_\ph^{[\mu}S_{1\perp}^{\nu]}}{Q}
             \,\bm a\!\cdot\!\bm S_{2\perp}\,\Bigg[
            {M_1}{M_2} D_{1T}^{\perp(1)}\frac{\overline G_T}{z_2}
          + {M_1}{M_2} \frac{\tilde D_T}{z_1}\overline G_{1T}^{(1)}
          - {m}{M_1}D_{1T}^{\perp(1)}\overline H_1
          - {M_2}^2H_1\frac{\overline E_T^{(1)}}{z_2}
          \Bigg]
\nonumber\\ &&
     {}+ i2\frac{\hat z_\ph^{[\mu}S_{2\perp}^{\nu]}}{Q}
             \,\bm a\!\cdot\!\bm S_{1\perp}\,\Bigg[
          - {M_1}{M_2} G_{1T}^{(1)}\frac{\overline D_T}{z_2}
          - {M_1}{M_2} \frac{\tilde G_T}{z_1}\overline D_{1T}^{\perp(1)}
          + {M_1}^2\frac{\tilde E_T}{z_1}(1)\overline H_1
          \Bigg]
\nonumber\\ &&
     {}- i2\frac{\hat z_\ph^{[\mu}a_\ph^{\nu]}}{Q}  \Bigg[
            \lambda_1\lambda_2\left(
                  {M_2}^2G_1\frac{\overline D_L^{\perp(1)}}{z_2}
                - {M_1}^2\frac{\tilde D_L^{\perp(1)}}{z_1}\overline G_1
            \right)
          - {M_1}{M_2} \,\bm S_{1\perp}\!\cdot\!\bm S_{2\perp}\,\left(
                  \frac{\tilde G_T}{z_1}\overline D_{1T}^{\perp(1)}
                - D_{1T}^{\perp}(1)\frac{\overline G_T}{z_2}
            \right)
\nonumber\\ && \hspace{21mm}
          - {m}{M_1}\left(
                 \,\bm S_{1\perp}\!\cdot\!\bm S_{2\perp}\,
                 D_{1T}^{\perp(1)}\overline H_1
               + H_1^{\perp(1)}\overline D_1
            \right)
          + {M_1}{M_2} \left(
                 H_1^{\perp(1)}\frac{\overline E}{z_2}
               - \frac{\tilde E}{z_1}\overline H_1^{\perp(1)}
            \right)
\nonumber\\ && \hspace{21mm} 
          + \lambda_1\lambda_2{M_1}{M_2} \left(
                 H_{1L}^{\perp(1)}\frac{\overline E_L}{z_2}
               - \frac{\tilde E_L}{z_1}\overline H_{1L}^{\perp(1)}
            \right) 
          \Bigg]
\nonumber\\ &&
     {}- i2\frac{\hat z_\ph^{[\mu}\epsilon_\perp^{\nu]\rho}a_\rho}{Q}  \Bigg[
            \lambda_1{M_2}^2G_1\frac{\overline D^{\perp(1)}}{z_2}
          + \lambda_2{M_1}^2\frac{\tilde D^{\perp(1)}}{z_1}\overline G_1
          - \lambda_1{M_1}^2\frac{\tilde G_L^{\perp(1)}}{z_1}\overline D_1
          - \lambda_2{M_2}^2D_1\frac{\overline G_L^{\perp(1)}}{z_2}
\nonumber\\ && \hspace{26mm} 
          + {M_1}{M_2} \left(
               \lambda_1 \frac{\tilde E_L}{z_1}\overline H_1^{\perp(1)}
             + \lambda_2 H_1^{\perp(1)}\frac{\overline E_L}{z_2}
             - \lambda_1 H_{1L}^{\perp(1)}\frac{\overline E}{z_2}
             - \lambda_2 \frac{\tilde E}{z_1}\overline H_{1L}^{\perp(1)}
            \right)
          \Bigg]
\Bigg\}.
\end{eqnarray} 

\pagebreak[4]

\section{Integrated twice-weighted hadron tensor}

We display only the leading terms of the hadron tensor weighted with
two factors $\left(\,\bm q_T^{}\!\cdot\!\bm a\,\right)\;
\left(\,\bm q_T^{}\!\cdot\!\bm b\,\right)$ and integrated
over $\bm q_T^{}$.  

\begin{eqnarray} 
\lefteqn{
\int d^2{\bm q_T^{}}\;
\left(\,\bm q_T^{}\!\cdot\!\bm a\,\right)\;
\left(\,\bm q_T^{}\!\cdot\!\bm b\,\right)\;
{\cal W}^{\mu\nu}_S=12 e^2 z_2 z_1\times\Bigg\{ }
\nonumber\\ &&
       + g_\perp^{\mu\nu} \Bigg[
           - \,\bm a\!\cdot\!\bm b\,
            \left({M_1}^2 D_1^{(1)}\overline D_1 
                + {M_2}^2 D_1\overline D_1^{(1)} 
                -\lambda_1\lambda_2{M_1}^2 G_1^{(1)}\overline G_1
                -\lambda_1\lambda_2{M_2}^2 G_1\overline G_1^{(1)} 
            \right)
\nonumber\\ && \hspace{14mm}
       +  2\,\bm S_{1\perp}\!\cdot\!\bm S_{2\perp}\,\,\bm a\!\cdot\!\bm b\,
            {M_1}{M_2} D_{1T}^{\perp(1)}\overline D_{1T}^{\perp(1)} 
\nonumber\\ && \hspace{14mm}
       +   \left(\,\bm a\!\cdot\!\bm S_{1\perp}\,
                 \,\bm b\!\cdot\!\bm S_{2\perp}\,
               + \,\bm a\!\cdot\!\bm S_{2\perp}\,
                 \,\bm b\!\cdot\!\bm S_{1\perp}\,\right) 
                {M_1}{M_2} 
           \left(G_{1T}^{(1)}\overline G_{1T}^{(1)}
                -D_{1T}^{\perp(1)}\overline D_{1T}^{\perp(1)} \right)
          \Bigg]
\nonumber\\ &&
          - \,\bm a\!\cdot\!\bm b\,
            \left(S_{1\perp}^{\{\mu}S_{2\perp}^{\nu\}}
                  +g_\perp^{\mu\nu}\,\bm S_{1\perp}\!\cdot\!\bm S_{2\perp}\,
            \right)
               \left({M_1}^2 H_1^{(1)}\overline H_1
                    +{M_2}^2 H_1\overline H_1^{(1)} 
                    -\frac{{M_1}^2}{2}H_{1T}^{\perp(2)}\overline H_1 
                    -\frac{{M_2}^2}{2}H_1\overline H_{1T}^{\perp(2)}
               \right)
\nonumber\\ &&
        -\left[
         \,\bm a\!\cdot\!\bm S_{1\perp}\,
         \left(S_{2\perp}^{\{\mu}b_\ph^{\nu\}}
              +g_\perp^{\mu\nu}\,\bm b\!\cdot\!\bm S_{2\perp}\,\right)
        +\,\bm b\!\cdot\!\bm S_{1\perp}\,
         \left(S_{2\perp}^{\{\mu}a_\ph^{\nu\}}
              +g_\perp^{\mu\nu}\,\bm a\!\cdot\!\bm S_{2\perp}\,\right)
         \right] 
             \frac{{M_1}^2}{2} H_{1T}^{\perp(2)}\overline H_1 
\nonumber\\ &&
        -\left[
         \,\bm a\!\cdot\!\bm S_{2\perp}\,
         \left(S_{1\perp}^{\{\mu}b_\ph^{\nu\}}
              +g_\perp^{\mu\nu}\,\bm b\!\cdot\!\bm S_{1\perp}\,\right)
        +\,\bm b\!\cdot\!\bm S_{2\perp}\,
         \left(S_{1\perp}^{\{\mu}a_\ph^{\nu\}}
              +g_\perp^{\mu\nu}\,\bm a\!\cdot\!\bm S_{1\perp}\,\right)
         \right] 
             \frac{{M_2}^2}{2}H_1\overline H_{1T}^{\perp(2)}
\nonumber\\ &&
       - 2{M_1}{M_2}\left(a^{\{\mu}b^{\nu\}}
                          +g_\perp^{\mu\nu}\,\bm a\!\cdot\!\bm b\,\right)
               \left( H_1^{\perp(1)}\overline H_1^{\perp(1)}
                     +\lambda_2\lambda_1 
                      H_{1L}^{\perp(1)}\overline H_{1L}^{\perp(1)}\right)
\nonumber\\ &&
       +   {M_1}{M_2}\left(a_\ph^{\{\mu}\epsilon_\perp^{\nu\}\rho}b_\rho
                          +b_\ph^{\{\mu}\epsilon_\perp^{\nu\}\rho}a_\rho\right)
               \left(\lambda_1 H_{1L}^{\perp(1)}\overline H_1^{\perp(1)}
                   - \lambda_2 H_1^{\perp(1)}\overline H_{1L}^{\perp(1)} 
\right)
\Bigg\}
\end{eqnarray} 
and
\begin{eqnarray} 
\lefteqn{
\int d^2{\bm q_T^{}}\;
\left(\,\bm q_T^{}\!\cdot\!\bm a\,\right)\;
\left(\,\bm q_T^{}\!\cdot\!\bm b\,\right)\;
{\cal W}^{\mu\nu}_A=12 e^2 z_1z_2\times\Bigg\{ }
\nonumber\\ &&
     {}- i\epsilon_\perp^{\mu\nu} \,\bm a\!\cdot\!\bm b\, \Bigg[
           \lambda_2{M_2}^2 D_1\overline G_1^{(1)}
          + \lambda_2{M_1}^2 D_1^{(1)}\overline G_1
          - \lambda_1{M_2}^2 G_1\overline D_1^{(1)}
          - \lambda_1{M_1}^2 G_1^{(1)}\overline D_1
          \Bigg]
\nonumber\\ &&
     {}+ i\left(S_{1\perp}^{[\mu}a_\ph^{\nu]}\,\bm b\!\cdot\!\bm S_{2\perp}\,
               +S_{1\perp}^{[\mu}b_\ph^{\nu]}\,\bm a\!\cdot\!\bm S_{2\perp}\,
          \right)
              {M_1}{M_2} D_{1T}^{\perp(1)} \overline G_{1T}^{(1)}
\nonumber\\ &&
     {}+ i\left(S_{2\perp}^{[\mu}a_\ph^{\nu]}\,\bm b\!\cdot\!\bm S_{1\perp}\,
               +S_{2\perp}^{[\mu}b_\ph^{\nu]}\,\bm a\!\cdot\!\bm S_{1\perp}\,
          \right)
              {M_1}{M_2} G_{1T}^{(1)}\overline D_{1T}^{\perp(1)}
\Bigg\}.
\end{eqnarray}

\newpage

\end{document}